\documentclass[10pt,preprint]{aastex}
\usepackage{lscape}

\slugcomment{ApJ, submitted}

\shorttitle{Lynx supercluster}
\shortauthors{Mei et al.}

\begin{document}


\title{Early-type galaxies at z = 1.3. I. The Lynx supercluster:
  cluster and groups at z=1.3. Morphology and color--magnitude relation}


\author{
Simona Mei\altaffilmark{1,2,3},
S. Adam Stanford\altaffilmark{4,5},
Brad P. Holden\altaffilmark{6},
Anand Raichoor\altaffilmark{1,7}, 
Marc Postman\altaffilmark{8},
Fumiaki Nakata\altaffilmark{9},
Alexis Finoguenov\altaffilmark{10},
Holland C. Ford\altaffilmark{11},
Garth D. Illingworth\altaffilmark{6},
Tadayuki Kodama\altaffilmark{9,12},
Piero Rosati\altaffilmark{13}
Masayuki Tanaka\altaffilmark{14},
Marc Huertas-Company\altaffilmark{1,2},
Alessandro Rettura\altaffilmark{4,11,15},
Francesco Shankar\altaffilmark{16},
E. Rodrigo Carrasco \altaffilmark{17},
Ricardo Demarco\altaffilmark{18},
Peter Eisenhardt\altaffilmark{19},
Myungkook J. Jee\altaffilmark{4},
Yusei Koyama\altaffilmark{14},
Richard L. White\altaffilmark{8}}
\altaffiltext{1}{GEPI, Observatoire de Paris, Section de Meudon, 5
  Place J. Janssen, 92190 Meudon Cedex, France}
\altaffiltext{2}{Universit\'{e} Paris Denis Diderot, 75205 Paris Cedex
  13, France}
\altaffiltext{3}{California Institute of Technology, Pasadena, CA 91125, USA}
\altaffiltext{4}{Department of Physics, University of California, Davis, CA 95616, USA}
\altaffiltext{5}{Institute of Geophysics and Planetary Physics, Lawrence Livermore National Laboratory, Livermore, CA 94551, USA}
\altaffiltext{6}{ UCO/Lick Observatories, University of California,  Santa Cruz 95065, USA}
\altaffiltext{7}{ INAF - Osservatorio Astronomico di Brera, via Brera
  28, 20121 Milano, Italy}
\altaffiltext{8}{Space Telescope Science Institute, Baltimore, MD,
  USA}
\altaffiltext{9}{Subaru Telescope, National Astronomical Observatory
  of Japan, 650 North A{'}ohoku Place, Hilo, HI 96720, USA}
\altaffiltext{10} {Max-Planck-Instit\"{u}t f\"{u}r  Extraterrestrische Physik, Giessenbachstrasse, D-85478 Garching, Germany}
\altaffiltext{11}{Department of Physics and Astronomy, Johns Hopkins
  University, Baltimore, MD 21218, USA}
\altaffiltext{12}{National Astronomical Observatory of Japan, Mitaka, Tokyo 181-8588, Japan}
\altaffiltext{13}{European South Observatory, Karl-Schwarzschild
  -Str. 2, D-85748, Garching bei Munchen, Germany}
\altaffiltext{14}{Institute for the Physics and Mathematics of the
  Universe, The University of Tokyo, 5-1-5 Kashiwanoha,
  Kashiwa-shi,Chiba 277-8583, Japan}
\altaffiltext{15}{Department of Physics and Astronomy, University of
  California, Riverside, CA 92521, USA}
\altaffiltext{16} {Max-Planck-Instit\"{u}t f\"{u}r Astrophysik,  Karl-Schwarzschild-Str. 1, D-85748, Garching, Germany}
\altaffiltext{17} {Gemini Observatory, Southern Operations Center,
  Casilla 603, La Serena, Chile}
\altaffiltext{18}{Department of Astronomy, Universidad de Concepci\'{o}n, Casilla 160-C, Concepci\'{o}n, Chile}
\altaffiltext{19}{Jet Propulsion Laboratory, California Institute of Technology, MS 169-327, 4800 Oak Grove Drive, Pasadena, CA 91109, USA}


\begin{abstract}
We confirm the detection of three groups in the Lynx supercluster, at $z\approx 1.3$, through spectroscopic follow--up and 
X--ray imaging, and we give estimates for their redshifts and masses. 

We study the properties of the group galaxies as compared to the two
central clusters, RX~J0849+4452 and  RX~J0848+4453. Using spectroscopic follow-up and multi-wavelength photometric
redshifts, we select 89 galaxies in the clusters, of which 41 are spectroscopically confirmed, and 74 galaxies in the
groups, of which 25 are spectroscopically confirmed.

We morphologically classify galaxies by visual inspection, noting that
our early--type galaxy (ETG)
sample would have been contaminated at the 30\% --40\% level by
simple automated classification methods (e.g. based on Sersic index).

In luminosity selected samples, both clusters and groups show high fractions of bulge--dominated
galaxies with a diffuse component that we visually identified as
a disk,  and which we classified as bulge--dominated spirals, e.g. Sas. The ETG fractions never rise above $\approx 50\%$ in

the clusters, which is low compared to the fractions observed in other massive clusters at $z\approx1$.  In the
groups, ETG fractions never exceed $\approx 25\%$. However, overall
bulge--dominated galaxy fractions (ETG plus Sas) are similar
to those observed for ETGs in clusters at
$z\sim1$. Bulge--dominated galaxies visually classified as
spirals might also be ETGs with tidal
features or merger remnants. They are mainly red and passive,
and span a large range in luminosity. Their star formation seems to have been quenched before experiencing a morphological
transformation. Because their fraction is smaller at
lower redshifts, they might be the spiral population that evolves into ETGs.

For mass--selected samples of galaxies with masses  $M>10^{10.6} M_{\sun}$ within  $\Sigma
> 500 Mpc^{-2}$, the ETG and overall bulge--dominated galaxy fractions 
show no significant evolution with respect to local clusters,
suggesting that morphological
transformations might occur at lower masses and densities. 

The ETG mass--size relation shows 
evolution towards smaller sizes at higher redshift in both clusters
and groups, while the late--type mass--size relation matches that observed locally. 

When compared to the
clusters,the group ETG red
sequence, shows lower zero points (at $\sim 2 \sigma$), and larger
scatters, both expected to be an indication of a younger galaxy
population. However we show  that any allowed difference between the
age in groups and clusters would be small when compared to the differences in age in galaxies of
different masses.

\end{abstract}


\keywords{galaxies: clusters --
          galaxies: elliptical and lenticular ---
          galaxies: evolution}

\def\hst{{\it HST}}



\section{Introduction}

The currently favored
cosmological model,  $\Lambda$CDM (e.g. Komatsu et al. 2011), predicts that some of the galaxies in clusters
formed at the peaks of dark matter fluctuations (a pristine population) while others were
accreted from surrounding infalling groups and the field
(infalling population). During this accretion process, we expect galaxy interactions with the environment
to quench star formation and trigger a stellar population (from active to passive)
and morphological (from late to early--type) transformation
(e.g. Diaferio et al. 2001; De Lucia et al. 2006; 
Poggianti et al. 2006, 2010; Faber et al. 2007; Moran et al. 2007).  Various transformation mechanisms have been proposed, some
driven by the galaxy environment -- including merging, galaxy harassment, and gas
stripping -- some related to intrinsic galaxy properties, such as mass, and still others to both (for
a review see Boselli \& Gavazzi 2006; for a critical summary of transformation
mechanisms in clusters:  Treu et al. 2003). 

The primary mechanisms driving galaxy
transformations in clusters arise from galaxy--galaxy and galaxy--cluster (mainly
tidal) interactions, dynamical friction, and interaction with the hot cluster
medium (ram pressure stripping, strangulation).  Dynamical friction drags galaxies towards the
cluster center where they can merge with the central galaxy.  For
galaxy--galaxy mergers to occur, however, 
the internal velocity dispersion of the galaxies must be greater than their
orbital energy. In galaxy clusters, where cluster velocity
dispersions are high, galaxy mergers are therefore unlikely, while
groups present an ideal environment for such interactions.

$\Lambda$CDM  semianalytical models (e.g. McGee et al., 2009) predict 
that $\approx~40\%$ of galaxies in a Coma-sized cluster at z=0.5 are
accreted from groups with masses larger than $M=10^{13} M_{\sun}$. At
higher redshift, the number of galaxies accreted by a cluster and the fraction of galaxies that suffered a transformation by the
environment are both lower. This model implies that at $z\approx 1.5$,
environmental effects should not yet be significant, but will be predominant only
at $z<1.5$. 

Moreover, both star formation and morphological transformations are observed to
occur mainly at redshifts $z < 1$ when star formation decreases and the early--type 
population increases. While the elliptical fractions show little
evolution, this increase is due to a lack of S0 galaxies at higher redshift (e.g. Postman et al. 2005;
Poggianti et al. 2006, 2010; Moran et al. 2007).

Both models and observations agree that at high redshift, e.g. $z>1$, we should observe galaxies before or
during these transformations and structures in the very first
phases of their assembly. As a result, we can observe galaxies in groups before their infall into the denser cluster
environment and disentangle galaxy transformations occurring
in different environments.

Recent work (e.g. Poggianti et al. 2009; Just et
al. 2010) has shown
that the evolution of the S0 galaxy fraction is stronger in poor clusters and groups compared to
more massive clusters over the redshift range $0<z<0.5$, suggesting that
transformation mechanisms associated with these environments (e.g.
galaxy-galaxy interactions as opposed to galaxy-cluster
interactions) dominate the S0 transformations.

At higher redshifts ($0.4<z<1.2$), however, Capak et al. (2007), have shown that
early--type galaxy fractions change faster in high density environments. Since
this is the range in redshift where most of the group accretion
takes place in $\Lambda CDM$, we would like to understand which 
galaxies are accreted by  galaxy clusters at this epoch. Their morphology, and
stellar population before accretion should provide clues to understanding
how the accretion process changes the high-density population. 
With this aim, we have studied in detail the cluster and group population of a supercluster 
at $z\approx1.3$, the Lynx supercluster.

Superclusters are the largest structures observed in the
Universe and the ideal laboratory for studying galaxy transformations
in different environments. Their dimensions can range from 10 to $\approx$100~Mpc,
and they are composed of two or more galaxy clusters and
surrounding groups. Studies of these large structures at high redshift have proliferated in
recent years (e.g. Gal et al. 2008; Lubin et al. 2009; Tanaka et
al. 2009), and reveal the different star formation rates and histories of galaxy populations
in different environments at $z \approx 1$.  Only in recent years have superclusters been discovered at
redshifts $z>1$. 

Thanks to recent wide--field multi--wavelength imaging with the
Suprime-Cam on the Subaru telescope, and infrared follow--up, the first supercluster at $z\approx
1.3$ was discovered by Nakata et al. (2005), the Lynx supercluster, followed by the discovery
of a supercluster around RDCS~J1252.9-2927 $z=1.24$ by Tanaka et al. (2007).
A recent paper by Tanaka et al. (2009) confirmed this last
superstructure through spectroscopic follow--up with FORS2 and GMOS on
Gemini-South, and presented detailed studies of the cluster and group galaxy star formation
activity.  Their results show a population of red
star-forming galaxies in the groups that are not found in the the central cluster. They interpret
their observed spectral line strengths and colors by the presence of
star formation heavily obscured by dust.  They suggest that this is
driven by galaxy--galaxy interactions that are more active in galaxy groups.

In this paper we study galaxy transformations in the Lynx supercluster at $z=1.26$, confirming
the detection of a supercluster with spectroscopic follow--up and
X-ray imaging.

The Lynx supercluster is composed of two clusters and numerous surrounding groups. The two
clusters, RX~J0849+4452 (the Eastern cluster; hereafter Lynx E) and
RX~J0848+4453 (the Western cluster; hereafter Lynx W) were detected in
the near--infrared (Lynx~W; Stanford et al. 1997) and through their
X--ray emission in the ROSAT Deep Cluster Survey  (Rosati et al. 1999), and spectroscopically confirmed respectively at z=1.261 (Rosati et al. 1999) and z=1.273 (Stanford et al. 1997). 

While Lynx E presents a more compact galaxy distribution, with a
central bright galaxy merger (van Dokkum et al. 2001, Yamada et
al. 2002) (that might eventually lead to a central  cD, cluster dominant galaxy), the galaxies in Lynx W are more sparsely distributed in a filamentary structure, and do not present an obvious central bright cD galaxy. Their X--ray emission as seen by Chandra (Stanford et al. 2001) follows the optical distribution, with  Lynx E showing a more compact spherical shape and Lynx W a more elongated one, and luminosities of $L_X^{bol}=(2.8 \pm 0.2) \times 10^{44}$~erg~$s^{-1}$ and $L_X^{bol}=(1.0 \pm 0.7) \times 10^{44}$~erg~$s^{-1}$, respectively (Rosati et al. 1999; Stanford et al. 2001; Ettori et al. 2004). 
Together with a more compact galaxy distribution, this is an indication that Lynx E is likely to be more dynamically evolved with respect to Lynx W. The velocity dispersion for Lynx W was measured at $\sigma = 650 \pm 170$~km/s (Stanford et al. 2001).
This value is consistent with estimates from the Jee et al. (2006) weak
lensing analysis of $\sigma = 740^{+113}_{-134}$~km/s and $\sigma = 762^{+113}_{-133}$~km/s for Lynx E and Lynx W, respectively. The most recent estimates of the cluster temperatures are $T=3.8^{+1.3}_{-0.7}$~keV and $T=1.7^{+1.3}_{-0.7}$~keV, respectively (Jee et al. 2006 measurements that are consistent with Stanford et al. 2001).

Deep, panoramic multi--color
($VRi',z'$--bands) imaging with Suprime--Cam on the Subaru telescope
identified seven galaxy groups around these two central clusters
(Nakata et al.\ 2005). This makes the Lynx region 
a unique laboratory, being one of the very few superclusters observed
at such a high redshift, and for this reason one of the best regions
at $z > 1$ in which we to study properties of evolving galaxies in
different environments within a structure that is in the process of assembly.

 This paper is the first of a series of papers centered on
the Lynx supercluster, and will be followed by an analysis of galaxy
masses and ages (Raichoor et al. 2011, Paper~II), their star formation histories as compared to the
field (Rettura et al. 2011; hereafter Paper~III), and the evolution of the
Kormendy and mass--size relations in different
environments (Raichoor et al. 2012, hereafter Paper~IV). In this paper, we will mainly concentrate on
galaxy morphology and the build--up of the color--magnitude relation. Our observations are presented in Section~2.  In Section~3 we
describe our sample selection. In Section~4 we describe our galaxy
morphology and color measurements. In Section~5, we present and discuss our results, and we conclude with Section~6.

We adopt the WMAP cosmology ($\Omega_m h^2 =0.1334$,
 $\Omega_{\Lambda} =0.734$, $h=0.71$; Komatsu et al. 2011).
All ACS filter magnitudes are given in the AB system (Oke \& Gunn
1983; Sirianni et al. 2005).

\section{Observations}

\subsection{Imaging and spectroscopy}

We have observed the main two clusters of the
Lynx superstructure as part of the ACS Intermediate Redshift Cluster Survey
(IRCS; Postman et al. 2005),  in the F775W (from hereafter $i_{775}$) and F850LP (from here after $z_{850}$) bandpasses with 
the ACS Wide Field Camera (WFC) in March and April 2004, for a total of 
7300 and 12200 secs, respectively.  The surrounding Lynx groups were
observed with the ACS WFC as part of an HST GO program (PI: Mei) between December 2005
and February 2006, in the same bandpasses, $i_{775}$ and $z_{850}$,
for a total of 6300 and 10500 secs, respectively. The ACS WFC
resolution (pixel size) is 0.05~\arcsec/pixel, and its  field of view is 202~\arcsec x 202~\arcsec. The ACS images were processed with the APSIS pipeline (Blakeslee et al. 2003b), with a {\it Lanczos3}  interpolation kernel.
Our photometry was calibrated to the AB system, with  synthetic
photometric zero-points of 25.678 and 24.867 mag, respectively, in $i_{775}$ and $z_{850}$  from the \textit{HST}/ACS website\footnote{http://www.stsci.edu/hst/acs/}.
A reddening of $E(B-V)=0.027$ was adopted (Schlegel et al. 1998), with $A_{i775}=0.054$ and 
$A_{z850}=0.040$.

 \textit{Spitzer}/IRAC (Fazio et al. 1998) $[$3.6$\mu$m$] $ and
 $[$4.5$\mu$m$] $~band imaging of the clusters was obtained in April
 2004 in 30 exposures of 200 s each for a total exposure time of 6000
 s. Lynx groups were observed in
 November 2005 and May 2006, in 12 exposures of 100 s each, for a
 total exposure time of 1200s (PI: S.A. Stanford).
The data were reduced using standard \textit{Spitzer} procedures. 
The basic calibrated data (hereafter BCD) frames were first corrected for muxbleed and pulldown using
custom IDL scripts (now available from the Spitzer Science Center),
and then processed with MOPEX to produce coadded mosaics.

\textit{Spitzer}/MIPS 24 $\mu$m were also used, from program 83 (PI Rieke; AORs 4758016
and 4758784) and were observed April 9, 2004 in the photometry
mode. To produce a mosaic, we combined the pointings towards the two
clusters in the larger Lynx supercluster.  The BCD frames were combined using the available {\tt MOPEX} software from
the Spitzer Science Center in 2006.  The processing required
eliminating the first two BCD frames in each exposure sequence, and
constructing an effective {\it flatfield} frame to remove some detector
artifacts.  A total of 224 BCD frames were combined into a single
mosaic with 2.45$\arcsec$ pixels.  We used the available mask files to
eliminate bad pixels.  The final mosaic had median background value of
35.21 MJy sr$^{-1}$ with a dispersion of 0.018 MJy sr$^{-1}$ or 2.53
$\mu$Jy per pixel.  The mosaic was aligned with the existing ACS
imaging using sources in common between the two mosaics.
Using tool {\tt APEX}, we constructed a source catalog of the mosaic.
The resulting catalog has a 5$\sigma$ flux limit of 68 $\mu$Jy at the
deepest part of the mosaic. However, because of source confusion and
the low effective exposure times in the rest of the data, the
effective flux limit of the catalog is  125 $\mu$Jy.  

The entire supercluster has been observed with Suprime-Cam at the
SUBARU telescope in the V, R, i', and z' bandpasses between November
2000 and March 2001. The Suprime-Cam has a field of view of 27 x 27
$arcmin^2$ with a resolution of 0.2 arcsec/pixel. Details can be found in Nakata et al. (2005). These observations
have been used to obtain galaxy photometric redshifts and to
identify seven groups around the two main clusters. We will
refer to Groups~1 through 7, with the same identifications as given in Nakata et al (2005).

To complement space and Suprime-Cam observations, we added to our analysis
ground--based observations in the  R, J, and 
Ks--band (PI: S. A. Stanford), described in our
companion paper Paper~II. Here, we briefly summarize
these observations. The two clusters were observed in the R--band with the red 
camera on the {\it Keck}/LRIS Spectrograph (Oke 1995
), for a total exposure time
of 8750 s.
The groups have \textit{Palomar}/COSMIC (Kells 1998) $R$--band
imaging, obtained in November 1999 with the COSMIC instrument , for
a total exposure time of 16200 s.  The near-infrared $J$ and
$K_s$ band imaging was obtained in December 2003 at the \textit{KPNO}
2.1~m telescope with the FLAMINGOS instrument (Elston 1998), for a
total exposure time of about 24000~s and 26700~ s, respectively.

The imaging observations were followed by spectroscopic campaigns
at the SUBARU, Keck, and Gemini telescopes.


Cluster spectroscopy was obtained from published data (9 members from Stanford et al. 1997,
8 from Rosati et al. 1999) and spectroscopic runs at the Keck
telescope with the  Low Resolution Imaging Spectrograph (LRIS; Oke et al. 1995) and
the Deep Imaging Multiobject Spectrograph (DEIMOS)  (for details see
Mei et al. 2006).  In total, we have 41 confirmed cluster members with
spectroscopic redshifts between 1.25 and 1.28.

Group spectroscopy was obtained from the Subaru, Keck and Gemini telescopes.
FN conducted a spectroscopic follow-up observation on January 2007
with FOCAS (Kashikawa et al. 2002) in MOS mode, at the Subaru telescope.
We used a 300 line $mm^{-1}$ grating blazed at 6500~\AA~
with the order-cut filter SO58. The wavelength coverage was between
5800~\AA~ and 10000~\AA~ with a pixel resolution of 1.34~\AA~$pixel^{-1}$. 
A slit width was set to $0''.8$, which gave a
resolution of $\lambda / \Delta \lambda \sim500$.
We selected four FOCAS fields which efficiently cover the large-scale
structure found by Nakata et al. (2005).
Target galaxies were primarily selected on the basis of their
photometric redshifts, as in Nakata et al.
%
%
Observing conditions were good with $\sim0.6-0.9$~arcsec seeing.
Total on-source exposures ranged between 8100~s and 14400~s.
Data reduction was performed by FN in a standard manner using {\sc IRAF}.
All the reduced 1d/2d spectra were visually inspected to estimate the redshift of each object. In this work we only
use spectroscopic redshifts with a good quality measurement.

Keck spectroscopy was performed with the instrument DEIMOS
by AS and BH on 18 and 19
October 2009 UT to cover Groups~1 and~2.
On 18 October, three exposures of 1800s each were obtained when conditions were clear with 0.7 arcsec seeing.   On 19 October four exposures of 
1800s were obtained when conditions were clear with 0.9 arcsec seeing.
The data were reduced by BH using the DEEP pipeline.  

We obtained 20 hours of spectroscopic observations with GMOS-N at the
Gemini North Telescope, between January and April 2006, on five
(Group~1, Group~2, Group~3, Group~6 and~7) of
the original seven groups observed for a total of $\approx$14400 secs each. These observations
were reduced by SM, RC and 
RD with the standard Gemini IRAF package released from the
Gemini Observatory.

From the spectroscopic redshifts obtained at SUBARU and Gemini, Groups~6 and 7
from Nakata et al. (2005) were confirmed to be structures at $z<1.26$.
Three groups, Groups~1,~2 and 3 were confirmed to be at the same
redshift as the two main clusters. The two clusters have a relative projected distance of
$\approx$~2~$Mpc$  and the two clusters and three spectroscopically
confirmed groups (see below) cover a scale of around 30~$Mpc^2$, in the WMAP cosmology
(Komatsu et al. 2011).

This paper will concentrate only
on those three spectroscopically confirmed groups, whose properties are summarized
in  Table~\ref{tab1} .


\subsection{XMM Newton imaging}

We retrieved XMM Newton imaging (Jansen et al. 2001; OBSID 0085150101, 0085150201, 0085150301) covering the Lynx supercluster
region to evaluate group detection and estimate masses. We use the data collected
by the European Photon Imaging Cameras (EPIC): the pn-CCD camera (Struder et al. 2001) and the MOS-CCD cameras
M1 and M2 (Turner et al. 2001), with total exposure time of 150ks
before cleaning, and net time of 70.0 ks, 89.8
ks, 90.7 ks, respectively, after cleaning (see below).

Initial data processing was performed using the XMMSAS version 7.1
(Watson et al. 2001; Kirsch et al. 2004; Saxton et al. 2005).
We then remove time intervals affected by solar flares to create the
calibrated event file, as in Zhang et al. (2004), and apply the four-stage background
subtraction of Finoguenov et al. (2007) to increase our capability to
detect extended emission. High background in some of the MOS chips (Kuntz \&
Snowden 2008) was checked and no instances were found. To create a final mosaic of cleaned images
in the area, we estimate the background in each image for each
instrument and correct for differences in sensitivity between instruments.

The residual map of extended sources is obtained by subtracting a PSF model using the
detected point sources from the background subtracted image and
smoothing the image with a Gaussian kernel of 32" width. The same
technique has been used to estimate the emission
in Tanaka, Finoguenov, Ueda (2010) and Gobat et al. (2011).

\begin{figure}
\centerline{\includegraphics[angle=0,scale=.60]{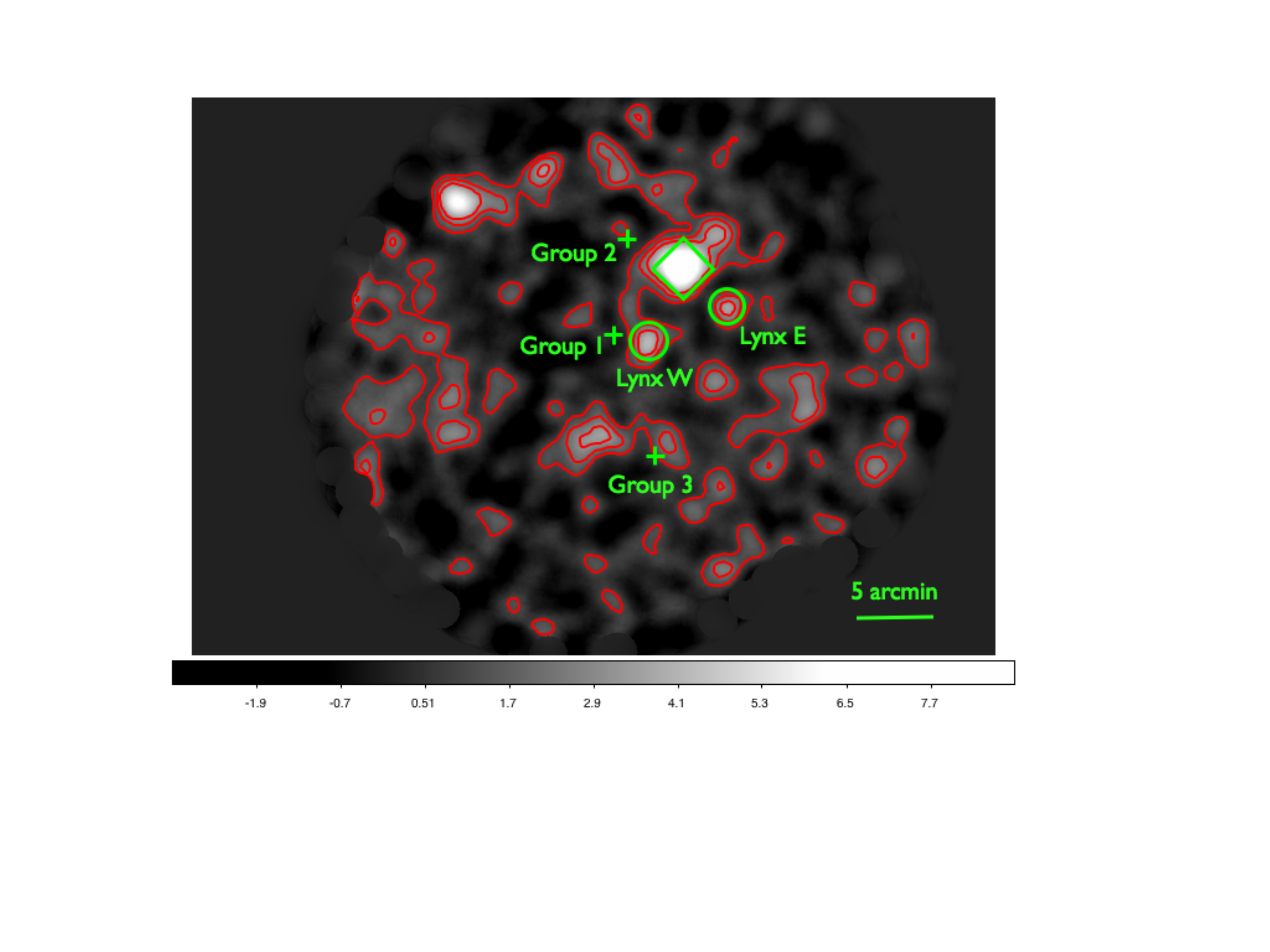}}
\caption{XMM Newton composite image of the Lynx supercluster
  ($\approx 50 \arcmin  \times 60 \arcmin$). The red
  lines show contours at 1, 2 and 3~$\sigma$. The two circles and the
  three crosses are the two clusters and three confirmed groups,
  respectively. Symbol sizes correspond to the estimated
  $R_{200}$. The diamond shows the known cluster at $z=0.57$ (Holden
  et al. 2001). The positions derived from the re--centered red galaxy
  overdensity do not exactly  match the X--ray peak emissions; however,
  we detect two of the groups (Group~2 and 3) and establish an upper
  limit on the X-ray emission
  for Group~1 (see text). The smoothing of the X--ray image introduces an
  uncertainty in astrometry of 32\arcsec. \label{xmm}}
\end{figure}

 Fig.~\ref{xmm} shows the XMM composite image. We detect Group~3 at 2.6~$\sigma$, Group~2 at
1.6~$\sigma$. Group~1 shows a 0.5~$\sigma$ excess, and there is an
excess of the X-ray emission (at around 2~$\sigma$) just close to it. From the X-ray
emission and scaling relations (Markevitch et al. 1998; see also
Bielby et al. 2010 for a detailed description of our X--ray analysis), we obtain a mass estimate for Groups~2 and 3 of $5.7 \pm
2.6  \times 10^{13} M_{\sun}$
and $5.0 \pm 1.7 \times 10^{13} M_{\sun}$, respectively, and a 95\% upper limit for the
mass of Group~1 of $M < 4.5 \times 10^{13} M_{\sun}$.

\section{Cluster and Group sample selection}

Our ACS images were aligned on the center of the cluster X--ray emission from
Stanford et al. (2001), for the two clusters, and on the center of the density overdensities as estimated from Nakata et
al. from a 10 nearest neighbor algorithm for the groups.  For this analysis, we 
selected cluster and group galaxies using newly estimated photometric
redshifts from our space and ground--based data, and a
Friend--of--Friends algorithm 
(FoF, Geller \& Huchra 1983; see also Postman
et al. 2005) with scale derived from the two clusters (see below).

As a first step in the galaxy selection, we consider as potential group members all galaxies
with SUBARU $Vri'z'$ photometric redshifts $0.8<z_{phot}<1.4$ from
Nakata et al. (2005), ACS magnitude brighter than  $z_{850}=24~mag$
(the limit magnitude 
for dependable visual morphological classification; see Postman et
al. 2005), and  ACS color $0.4<(i_{775}-z_{850})<1.4$. This first
selection is larger than the initial 
$1<z_{phot}<1.35$ range proposed by Nakata et al. (2005).  It is based on
the range in SUBARU photometric redshifts and ACS colors covered by
spectroscopically confirmed
early and late--type cluster members from our spectroscopic follow-up (see
below) and the large range of ACS colors predicted from Bruzual \&
Charlot (2003; BC03) stellar population
models for early and star--forming galaxies at $z=1.26$. With this
choice we obtain 195 objects with estimated completeness at more
than 90\% (from the spectroscopic follow--up, and the photometric redshift
uncertainties) at our limiting magnitude $z_{850}=24$~mag.

From our spectroscopic runs on the groups, we obtained 52 redshift measurements, of 
which 25 are spectroscopically confirmed members with
redshifts between 1.25 and 1.28, with average spectroscopic redshift
$z=1.262 \pm 0.007$ (Group~1; from 9 members), $z=1.260 \pm 0.006$
(Group~2; from 7 members), $z=1.263 \pm 0.005$ (Group~3; from 9
members). These values are slightly different from those published in
Paper~II because we have obtained new spectroscopically confirmed
members since publication.

Using our spectroscopic
redshifts, we then eliminated from this sample all
27 spectroscopically confirmed outliers (9 in Group~1, 11 in Group~2, 7
in Group~3).  The contamination rate in all group spectroscopic
samples, selected with SUBARU photometric redshifts, is $\approx 50\%$ and similar for the early and late--type (see
below for morphological classification) samples.

For the remaining 168 objects, we eliminated Sextractor multiple
identifications (in the ACS images) and derived a new set of photometric
redshifts using our space and ground--based multi--wavelength
photometry (in the range 0.6-4.5~$\mu m$) as described in our
companion paper Paper~II. We also re-estimated
photometric redshifts for the 94 cluster galaxies selected with the
same criteria in color and photometric redshift.

To estimate new photometric redshifts, we used the photometry from
Raichoor et al. (2011; see this paper for details on photometry and
the sky background subtraction) and the public software
Le~Phare (Arnouts et al. 2002; Ilbert et al. 2006), based on a 
$ \chi^2$ template fitting method. The best photometric redshift is
calculated using the median of the Probability Distribution Function.
For our input parameters, we
followed Ilbert et al. (2010):  BC03 templates, solar metallicity, an exponentially
decaying star formation with $\tau$ in the range 0.1 to 5~Gyr, and a
Calzetti et al. (2000) extinction law with E(B−V) in the range 0 to 0.5.

These new photometric redshifts include optical photometry, ACS $i_{775}$ and $z_{850}$, near--IR and intermediate IR
photometry from Spitzer/IRAC, and help us to better define the
statistics of our sample. In total we have 41 and 25 spectroscopically
confirmed members in the clusters and groups, respectively, for a total
of 66 spectroscopically confirmed members.
At z=1.3, photometric redshift uncertainties are larger than a galaxy group extension in redshift space (e.g. Evrard et al. 2008, George et al. 2011).
We use an empirical approach to estimate the completeness and purity
of our sample. Most (90\% of the entire sample; 95\% of the early--type, 83\% of the late--type ) of our
spectroscopically confirmed members show new photometric
redshifts in the range $0.92 <z_{phot}<1.36$. The remaining 10\%  show
$1.36<z_{phot}<2$. We use this new range in photometric redshift
(e.g. $0.92 <z_{phot}<1.36$) to
select our final sample. In this range, we also find only 12 group
outliers, making our estimated contamination level decrease from 50\% to
$\approx$~20\% with this new photometric redshift selection. 
From these empirical estimates, we consider our sample to be $\approx
95\%$ (for the early--type galaxies) and $\approx
83\%$ (for the late--type galaxies) complete and contaminated at $\approx
20\%$. 

To verify these estimates, we used as test sample the George et
al. (2011) COSMOS group and field sample, that uses Ilbert et
al. (2009) photometric redshifts obtained with SED fitting using 30
bandpasses and with the software Le~Phare with the same input parameters as
ours (once considering our adaptations to a z=1.26 sample). 
When using Ilbert et al. (2009) photometric redshifts as reference
redshifts (hereafter $z_{phot-COSMOS}$), calibrated on
VIMOS/VLT and DEIMOS/Keck spectroscopy, and estimating
photometric redshift for the COSMOS sample using only the four
bandpasses we use in this paper (hereafter $z_{phot-4b}$), we obtain a difference in photometric
redshifts ($z_{phot-4b}$-$z_{phot-COSMOS}$)
consistent with zero, with a standard deviation $\sigma = 0.12$.
Ilbert et al. (2009) estimates the uncertainty on their photometric
redshifts to be $\sigma_z \approx 0.03$, at $z=1.26$ and at the magnitude depth of
our sample. We estimate then the uncertainty on our photometric
redshift measurement to be $\sigma_{opz} = 0.12$. 
When we estimate how many COSMOS galaxies with $z_{phot-COSMOS}$ in the range 
$1.26 \pm 3\sigma_{opz}$ are recovered within the range $1.26 \pm
3\sigma_{opz}$  in $z_{phot-4b}$, we find $87\%$ of the objects. 
On the other hand, when estimating how many objects 
with $z_{phot-4b}$ in the range $1.26 \pm 3\sigma_{opz}$, also have
$z_{phot-COSMOS}$ in that range, we obtain $85\%$.
When instead of $1.26 \pm 3\sigma$, we use our empirical selection,
e.g. the range in redshift  $0.92 <z_{phot}<1.36$, these percentages
become $96\%$ and $87\%$ , very similar to the percentage of
completeness and purity, respectively, that we have estimated based on our empirical
analysis.
These value are also compatible with more sophisticated estimates of
uncertainties on
photometric redshift and group membership estimated in
George et al. (2011). 

Therefore, we estimate our photometric redshift
uncertainty to be $\sigma = 0.12$ at $z=1.26$, and that our
photometric redshift
selection is $\approx
95\%$ complete for the early--type galaxies ($\approx
87\%$ for the late--type), with a contamination of $\approx
20\%$.

To further select group members, we applied a FoF algorithm.
As starting center we initially used the
overdensity centers defined by Nakata et al. (2005). When galaxies
have morphological classifications (see below), we redefine our final group
centers as luminosity--weighted centers from our early--type red
population.
Our linking scale was normalized on the two
main clusters, and chosen as the scale at which we select all
spectroscopically confirmed cluster members. We obtain a linking scale
corresponding to a local distance of $0.54$~Mpc, normalized to z=1.26
and to our magnitude range as in Postman et al. (2005). 


We have also re-selected
the cluster galaxy sample exactly in the same way as the group
sample, using photometric redshifts and the
ACS GTO spectroscopic follow-up described in Mei et al. (2006,
2009). Our cluster sample remains unchanged.

Our final sample comprises 89 galaxies in the clusters, of which 41
are spectroscopically confirmed, and 74 galaxies in the
groups, of which 25 are spectroscopically confirmed.

\section{Morphological classification and photometry}

\subsection{Morphological classification}

\begin{figure}
\includegraphics[angle=0,scale=.50]{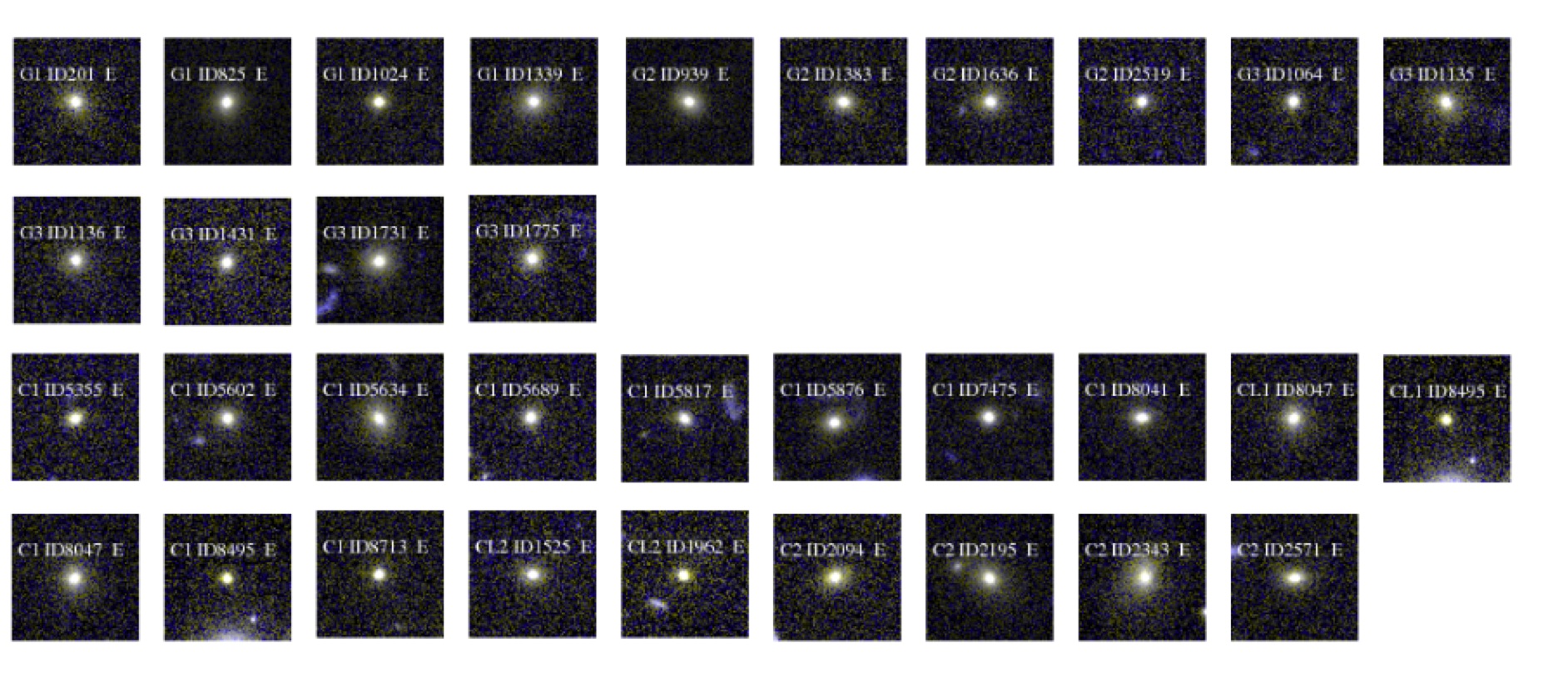}
\includegraphics[angle=0,scale=.50]{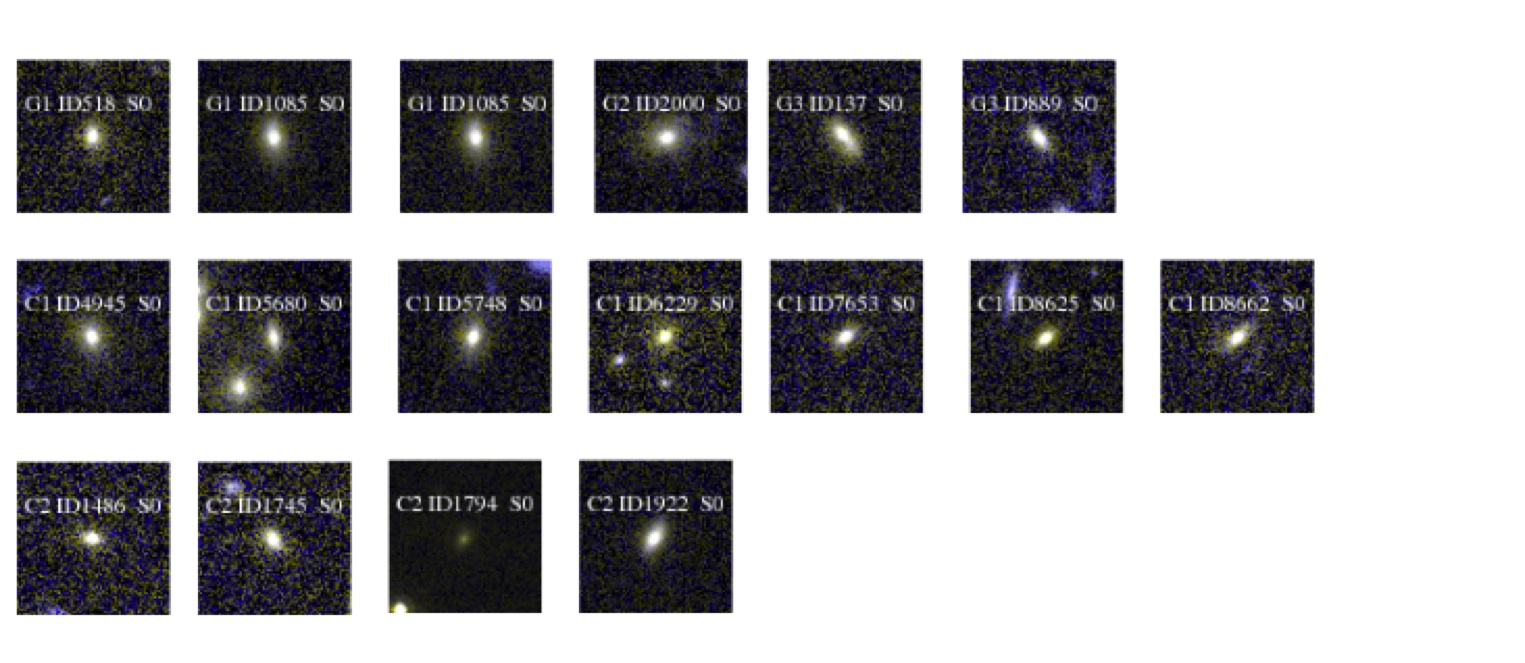}
\includegraphics[angle=0,scale=.50]{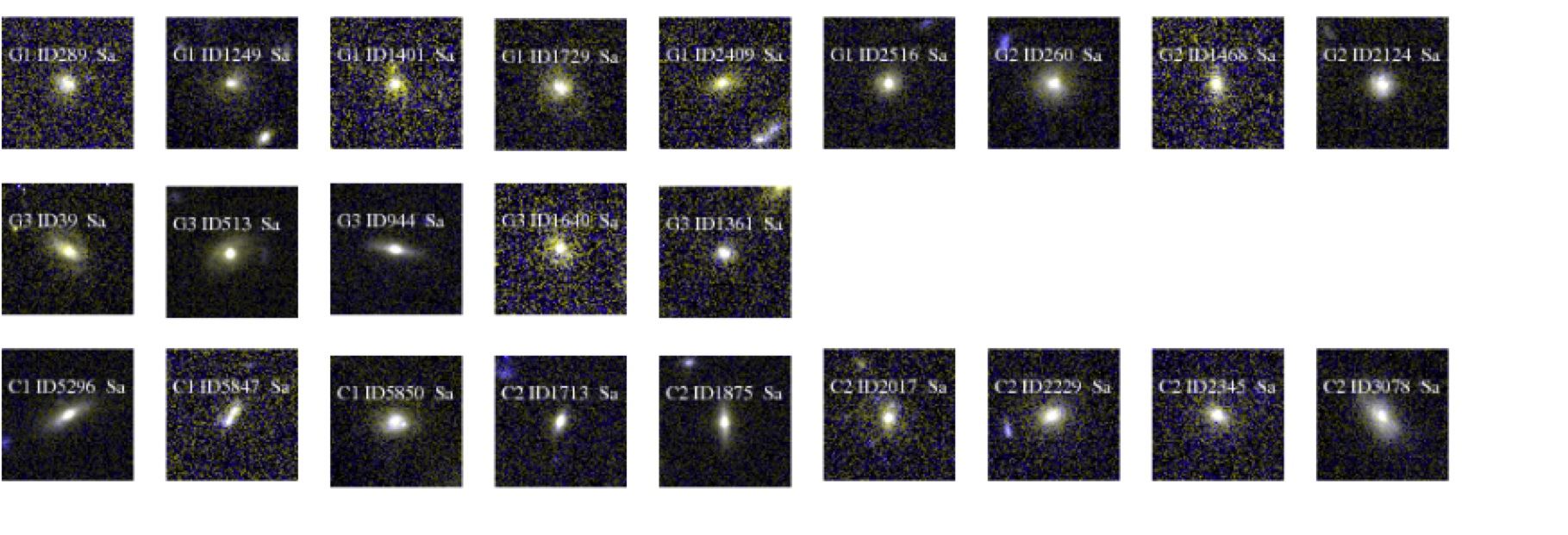}
\caption{Galaxies classified as elliptical, S0 and Sa (from top to bottom) in the cluster and group
  sample. The galaxies labeled as C1 and C2 belong to Lynx~E and~W,
  respectively. The galaxies labeled as G1, G2, G3 belong to Group~1,
  Group~2 and Group~3, respectively. \label{fig1}}
\end{figure}

\begin{figure}
\includegraphics[angle=0,scale=.50]{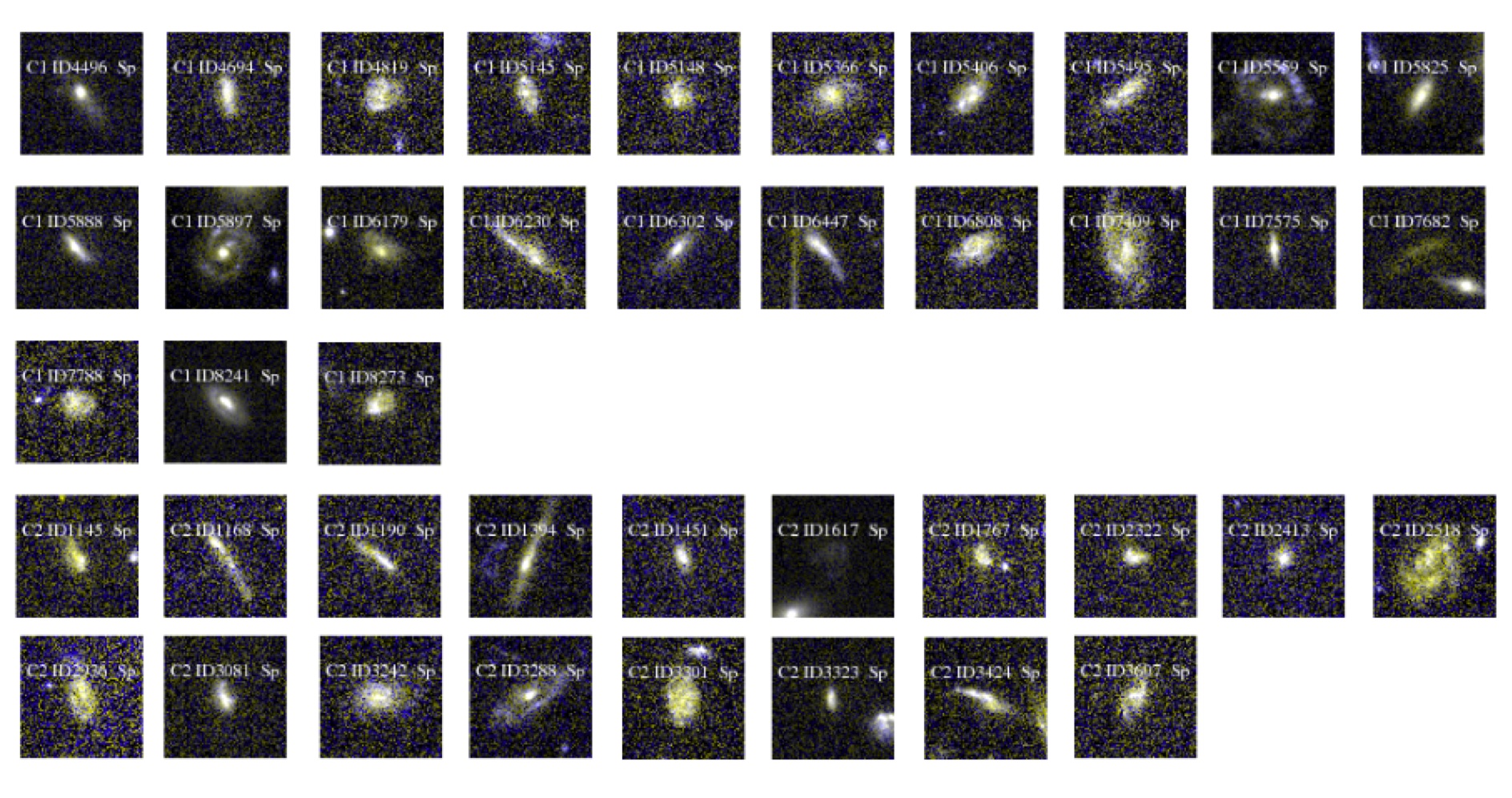}
\includegraphics[angle=0,scale=.50]{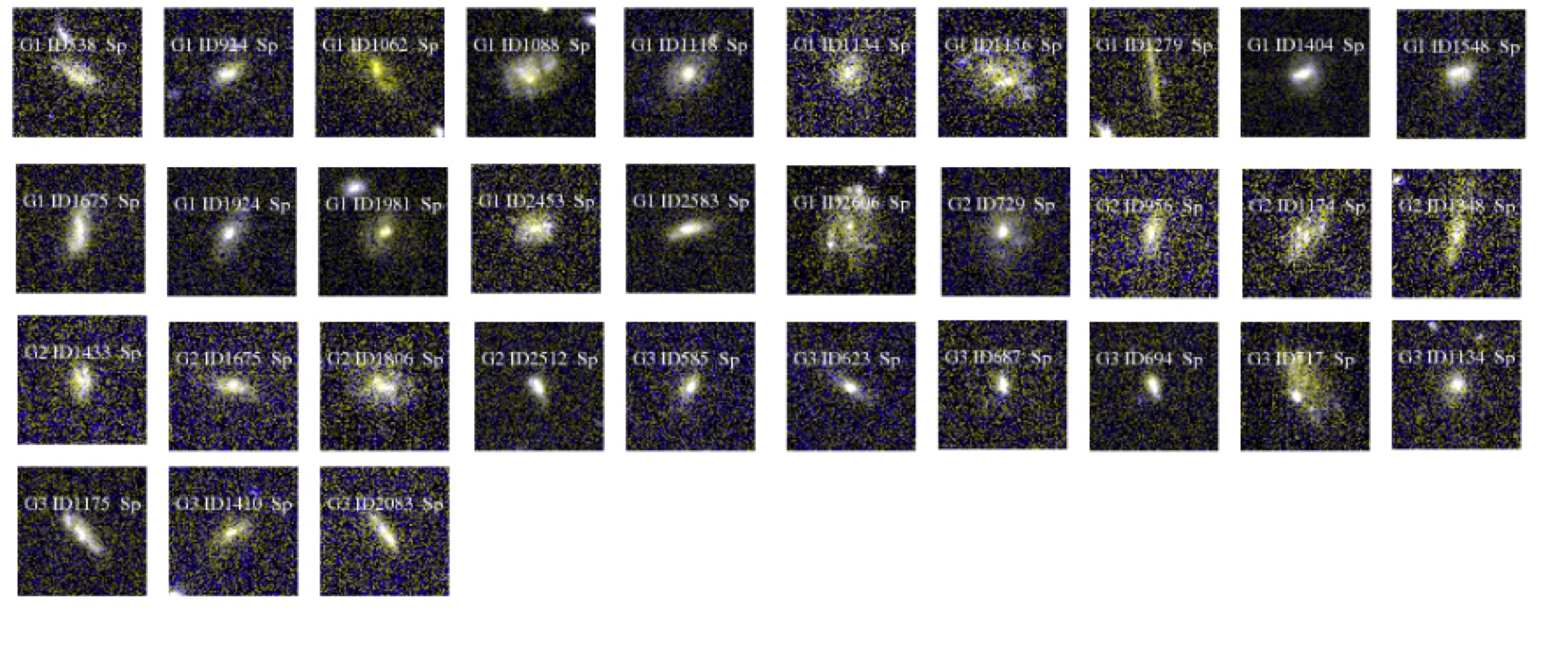}
\caption{Galaxies classified as later (Sb, Sc and Sd) spiral and
  irregular types in the cluster and group
  sample. The galaxies labeled as C1 and C2 belong to Lynx~E and~W,
  respectively. The galaxies labeled as G1, G2, G3 belong to Group~1,
  Group~2 and Group~3, respectively.  \label{fig2}}
\end{figure}

\begin{figure}
\includegraphics[angle=0,scale=.60]{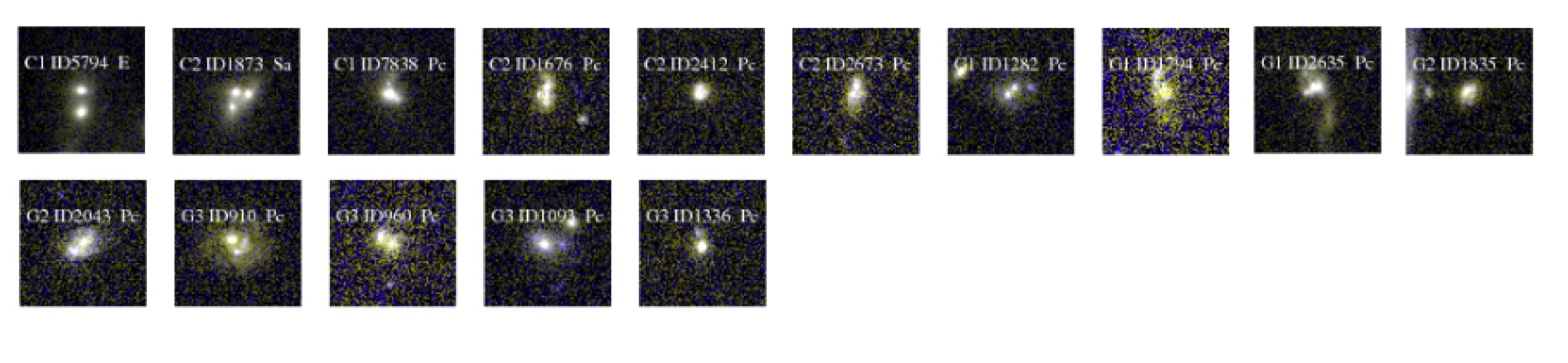}
\caption{Galaxies classified as mergers or peculiar in the cluster and group
  sample. The galaxies labeled as C1 and C2 belong to Lynx~E and~W,
  respectively. The galaxies labeled as G1, G2, G3 belong to Group~1,
  Group~2 and Group~3, respectively. \label{fig3}}
\end{figure}

All our selected group members
have been morphologically
classified  using the same visual classification as Postman et
al. (2005), based on the ACS $z_{850}$ images, the closest to rest
frame B--band (conventionally used for morphological classification) at $z=1.26$. Specifically, we classified galaxies as early-type
(ETGs, defined as elliptical and S0s), late-type (Sa, Sb, Sc), and
irregular/undefined. With respect to Postman et al. (2005), we have
added a new class of clumpy/uncertain objects, and objects presenting
interactions or visible tidal tails to study their occurrence in the
cluster and group population. We will refer to these objects as {\it disturbed}. Three of us, SM, MP and AS, have independently classified
all galaxies in this sample and compared our results. As already noted in other work (e.g. Postman et al
2005), our classification of early-type vs late-type galaxies was the
same in around $90\%$ of the cases, while it is the same in
approximately $70\%$ of the cases when distinguishing the early (e.g. elliptical from S0) and late-type subclasses. From
here on, the terms early--type and late--type galaxy refer,
respectively, to galaxies morphologically classified as elliptical and
S0, and as spiral galaxies according to this visual classification. 
Sa galaxies are classified so because they show diffuse features that
look like spiral arms around a massive bulge. We call this class of
galaxies {\it bulge--dominated spirals} (hereafter BDS or galaxies visual
classified as Sa), and we cannot exclude that
the diffuse component that we identify as a disk might be tidal structures or merger remnants.
We show our new visual classification in Figs.~\ref{fig1} to~\ref{fig3}.

\begin{figure}
\includegraphics[angle=90,scale=.50]{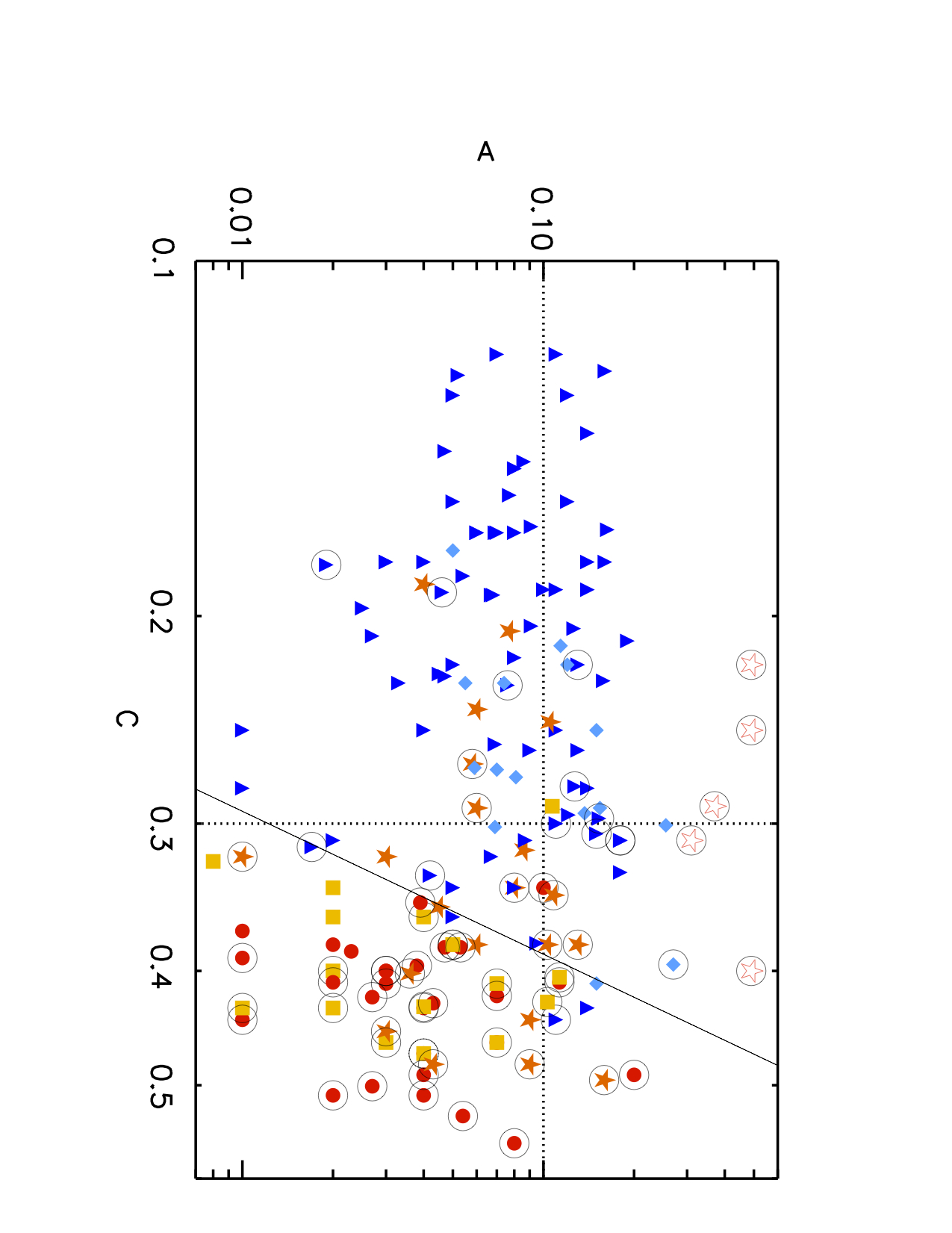}
\caption{Our visual morphological classification in the Concentration (C) vs
  Asymmetry (A) plane. Elliptical, S0, Sa and late Spirals are shown,
  respectively, as red circles, yellow squares, orange stars, and blue
  triangles. Clumpy/undefined and late--type merger (e.g. disturbed) galaxies are shown has light blue
 diamonds and bulge mergers as red open stars. Most of our
  early--type population has $C<0.3$ and $A<0.1$ (dotted lines; see the text for
  more details). If the galaxies were only 
  automatically classified, though, compact and symmetric galaxies ($C<0.3$
  and $A<0.1$) would include around half of the Sa population and
   $\approx$~10\% of the later type spirals, for a total
  contamination of the early--type sample of $\approx$~37\%.  The
  continuous line maximizes the recovery of the two separate early and late-type
  classes. When using this line to separate classes, we 
  classify as early--type 96\% of the visually classified early-type
  galaxies,  42\% of the Sa and 7\% of the late--type spirals. The
  distribution of the morphological types in this new ``early--type''
  region is visual early--type 73\%, Sa 23\%, late type 4\%.  Late-type galaxies would contaminate the
  early--type sample at $\approx$~30\%, mostly because Sa galaxies
  would be classified as early--type. Empty circles show galaxies
  with $n>2.5$, often used to separate early from late--type
  galaxies. Had we used this criteria to define our
  early--type sample, we would have suffered a contamination of $\sim
  40\%$, especially from Sa galaxies. \label{fig4}}
\end{figure}

For each of the galaxies, we also measured the 180$^{\circ}$ rotational asymmetry 
$A$ and image concentration $C$ parameters (Abraham et al. 1994; Conselice
et al. 2000) using the PyCA software (Menanteau et al. 2006). 
In Fig.~\ref{fig4}, we show our visual morphology in a
concentration/asymmetry plane.  Most of our visually classified  ETGs have
$C>0.3$ and $A<0.1$. In this region, we found 89\% of our total
early--type  population (excluding the bright early--type pair and
triplet - see below), 50\% of the
Sa, 13\% of the late spirals, irregulars or disturbed.

The distribution of the morphological types in this region is
66\% of visual early--type, 18\% of Sa galaxies, 
16\% of late--type spirals. We can
separate at ~10\% early from late--type galaxies (excluding Sa galaxies) just by simple cuts
in the C/A plane, and both samples will have more or less concentrated
Sa, around half among the highly concentrated/early-type dominated
region and the other half in the less concentrated/late--type
region. Sa galaxies, because of their large/concentrated bulges are then a
class that highly contaminates the ETGs. 

If we draw a line in the log(A) vs log(C) plane, following Abraham et
al. (1996) and Menanteau et al. (2006), and optimizing it for our sample, we obtain two regions that
maximize the recovery of two separate classes of early and late--type
galaxies (e.g. maximizing the
maximum percentage of early--type and the less of late--type
contamination in the early-type region ). 
When we use this line to separate classes, we 
  classify as early--type 96\% of the visually classified early-type
  galaxies,  45\% of the Sa and 8\% of the late--type spirals. The
  distribution of the morphological types in this new ``early--type''
  region is visual early--type 73\%, Sa 23\%, Late type 4\%. 
 When we exclude Sa galaxies, late-type galaxies are a small contamination for
 ETGs selected in this way. However, when the entire late--type
 population is considered, they would contaminate the
  ETG sample at $\approx$~30\%, mostly because of the Sa
  galaxies.
This classification would do better than the simple cut above, but we
will still have  $\approx$~30\% of the ``early--type'' that do
visually show a disk, and that we visually classify as Sa.

In Fig.~\ref{fig4}, we also show galaxies with Sersic index $n>2.5$
(see Sect.~4.2),
often used to separate early from late--type galaxies. In the combined cluster and group sample, we find that 70\% of the early-type galaxies have $n > 2.5$, and 78\% of the
late-type galaxies have  $n < 2.5$. Moreover, 57\% of the galaxies with $n>2.5$
are ETGs, 86\%  of the galaxies with $n<2.5$ are late-type. If we would
classify early--type galaxies objects with $n>2.5$, our ETG sample
would be contaminated at $\approx 40$\% by late--type galaxies.

Early--type galaxies identified only
by their compactness (or Sersic index) are highly contaminated by
BDS. Bernardi et al. (2009) obtain
 similar results in the local Universe with a detailed
analysis of the local Sloan Digital Sky Survey (SDSS; York et al. 2000) ETG sample. 

When using more advanced non--parametric techniques from Huertas-Company et al. (2008, 2009, 2011)~\footnote{the code and the training sets are available at http://gepicom04/galsvm.htm}, calibrated on the SDSS visual classifications, they are more efficient and we recover 89\% of our visual ETGs with 5\% contamination (2 galaxies which are visually classified as Sas). Interestingly, the algorithm manages also to isolate Sa galaxies with reasonably good accuracy: 6/9 visual Sas are classified by the algorithm as early-spirals, 1 as a late spiral and 2 of  as early-type.

In this paper, we only use our visual morphological
classification and discuss how some of our results would change based on
a simple automated classification.

Our first result is an absence of massive/bright red bulge
pairs or triplets (dry mergers) in the groups. While we do observe a double red bulge and a triple
red bulge merger in the two clusters (see also van Dokkum et al. 2001, Yamada et al.
2002, Mei et al. 2006b), no merger of this kind is visible in any of
the groups. 

Four disturbed objects are observed
in the two clusters (3 in the less evolved Lynx~W), and 9 were found in the groups (Fig~\ref{fig3}). 
Disturbed morphologies appear as blue compact galaxies with a
companion of the same mass (ID~1282) or smaller (ID~1093), clumpy
galaxies, with clumps of similar sizes (9 gal., see Fig~\ref{fig3}) or
with a central bulge and smaller clumps (ID~910, ID~2043), or as
clumps/bulges with large tidal tails (ID~2635, ID~1794).
This confirms that galaxies in groups are going through
transformations, driven in part by wet/blue late--type mergers, or
smaller companion. Most of them show clumpiness, with clumps
of similar sizes. These transformations are rarely seen in Lynx~W (1), while Lynx~E
shows three clumpy galaxies. 


These
observations indicate that wet merger activity is present both
clusters and groups and dry merger
activity is only present in the two clusters. Clearly though, since we are studying a
single supercluster,
 these results are not statistically significant (as in the case of other
 work at these high redshifts) and should be confirmed by larger statistical samples of superclusters of galaxies at these redshifts, 
unfortunately unavailable at this time (e.g. at present, this is the best we can
do).

\begin{figure}
\centerline{\includegraphics[angle=0,scale=.60]{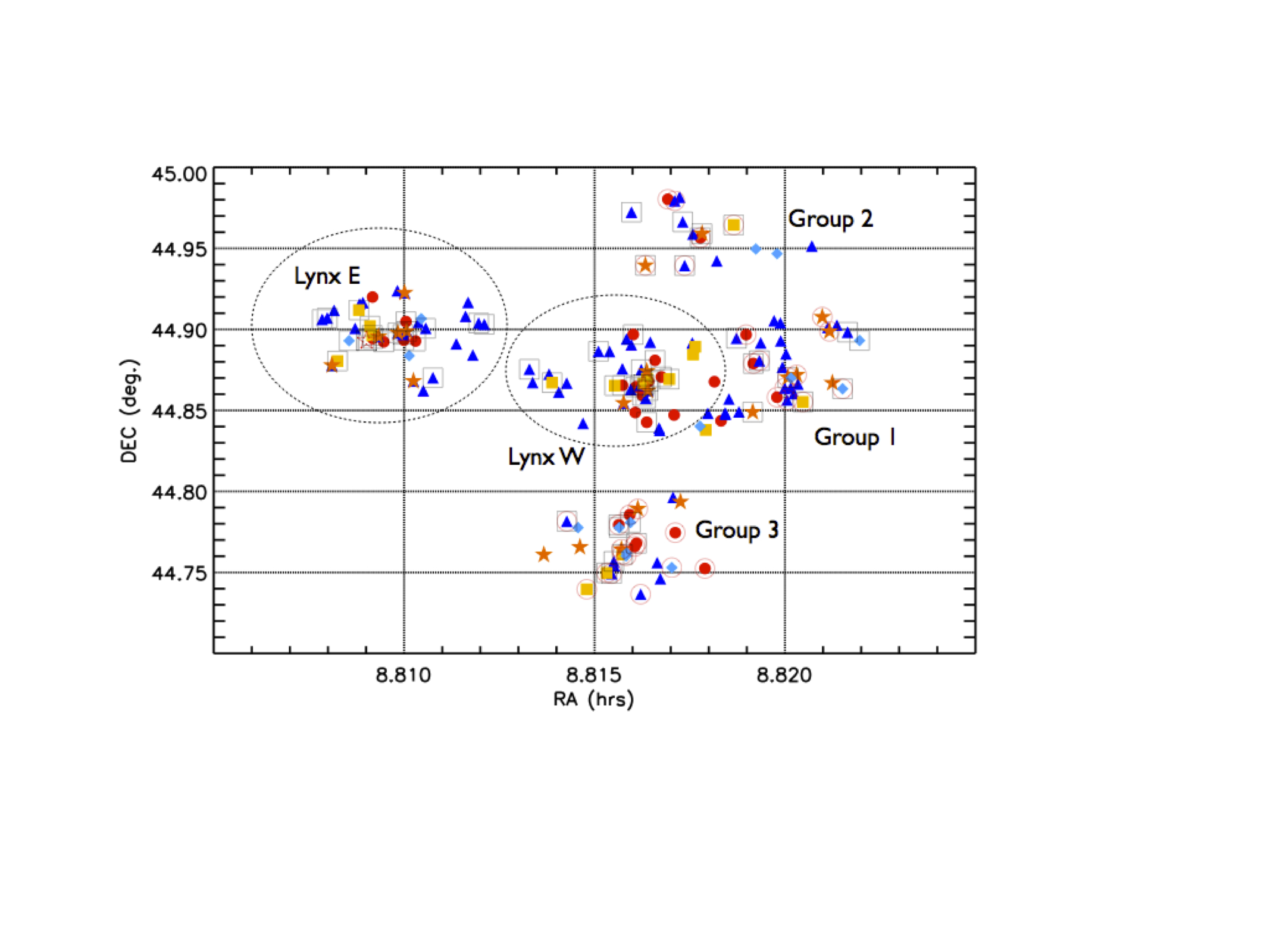}}
\caption{Spatial distribution of our  sample. Visual morphological
  classes are
  shown with the same symbols are as in Fig.~\ref{fig4}. 
Empty red circles flag galaxies with $0.8<(i_{775}-z_{850})<1.2$,
e.g., red galaxies. The red overdensities are clearly
identified. Empty black squares show spectroscopically confirmed
members. We show galaxies assigned to the two clusters within the
dotted lines. Group~1 is spatially connected (e.g. by the FoF algorithm)
to Lynx~W.  We consider this group as a separate structure, because its center lies at $\sim 1.1 R_{200}$ from the center of the
cluster and it extends to $\sim 2 R_{200}$, with an area of very low density between $0.5$ and 1~ $\sim
R_{200}$. Group~1 might be close to merging with Lynx~W, or in the merger process.\label{fig5}}
\end{figure}

In Fig~\ref{fig5}, we show the spatial distribution of our selected
sample, the morphological types and the red --
$0.8<(i_{775}-z_{850})<1.2$ --
overdensities in the clusters and the groups. Group~1 is very close to
Lynx~W, and its galaxies are connected by a FoF algorithm. We
consider them as two separate structures, nevertheless, because the center of
the group lies at $\sim 1.1 R_{200}$ from the center of the
cluster and extends to $\sim 2 R_{200}$, with an area of very low density between $0.5$ and 1~ $\sim
R_{200}$. It might be close to merging, or in the merger process. High
resolution X--ray imaging would give us more elements to understand
the nature of their interaction.

\subsection{Structural parameters, color and magnitude measurement}

\begin{figure}
\centerline{\includegraphics[angle=90,scale=.70]{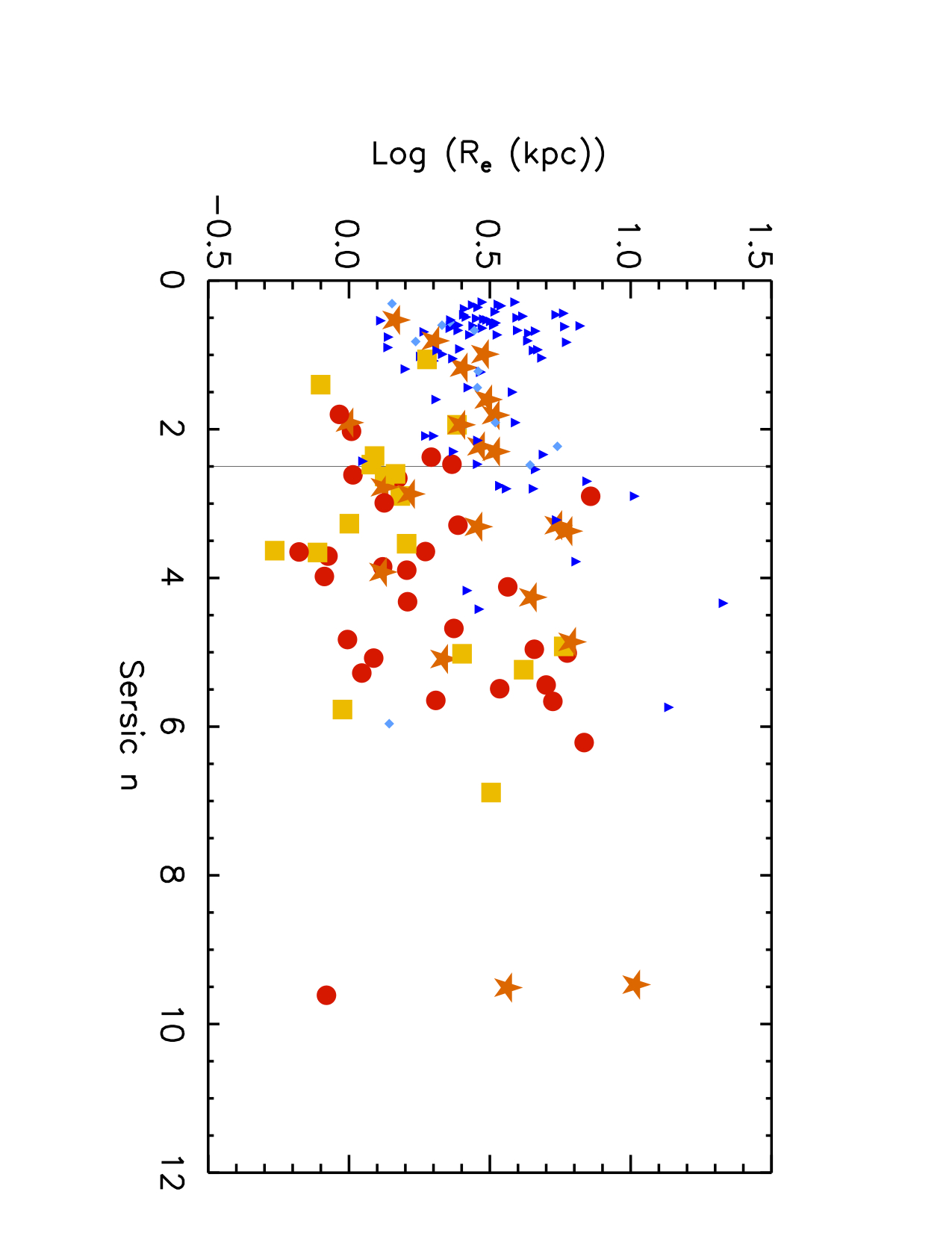}}
\caption{Size as a function of Sersic index $n$ for all
  morphological types. Symbols are the same as Fig.~\ref{fig4}. The
  continuous vertical line shows $n=2.5$ often used to separate early
  from late--type galaxies (see caption of Fig~\ref{fig4} for
  discussion). Most of our galaxies have $n<6$. In some cases, we
  find larger Sersic indexes, but we do not observe any indication of bias in
  the size estimation (e.g. we do not see a correlation between $n$ and
  size). Symbols for late--type galaxies are smaller because of the
  high density of this kind of galaxies at $n<1$. \label{sersic}}
\end{figure}

We derived galaxy structural parameters using GALFIT  (Peng et
al. 2002) in the ACS $z_{850}$ imaging,
fitting elliptical
Sersic models to each galaxy image. With respect to our previous work
(e.g. Mei et al. 2009),
in this work we did not constrain the values of the {\it Sersic} index for most of the
sample, as suggested in Peng et al. (2002). Only the five galaxies found in double or triplets in the clusters
required simultaneous fitting and a constraint $n <4$ in order for the fit
to converge. For each galaxy, we obtain its {\it ellipticity} (defined
as 1-$q$, where $q=\frac{b}{a}$ is the axial ratio),  its average
half-light radius $R_e$, and {\it Sersic} index $n$.  
In Fig.~\ref{sersic}, we plot size as a function of the Sersic index
$n$ for our different morphological types. While most galaxies have Sersic
index $n<6$, for some the fit converged to higher values of $n$. In
these latter cases, we do not see any evidence for a bias in size
estimation. To see if there is a bias in size estimation, e.g. a
correlation between $n$ and  $R_{e}$, we calculate Pearson
coefficients and the probability of correlation between these two
variables.  We obtained a probability of correlation $<60\%$
and a probability to have a random distribution $>95\%$, for both the entire sample and for the
early and late--type populations separately.

We measured galaxy colors following Mei et al. (2006a,b; see also
Blakeslee et al. 2006).  Our colors
were measured inside a circular aperture scaled by the galaxy average
half-light radius $R_e \sqrt{q}$ (van Dokkum et al. 1998, 2000; Scodeggio
2001), to avoid galaxy gradient effects.

For early-type galaxies, we have checked (Mei et al. 2006a) that our colors do not change (within
the uncertainties) if the  effective radii are calculated via a two component 
(Sersic bulge + exponential disk) surface brightness decomposition 
technique using GIM2D (Marleau \&  Simard 1998; Rettura et al. 2006) that better fits 
the galaxy light profiles. For late-type galaxies,
we used a single profile decomposition, so that their aperture radius would
be larger than their bulge, but this doesn't take into account the
extension of their disk.  For early-type spirals (Sas), this value is very close to
their bulge size.

We estimate uncertainties on the color by
adding the uncertainty due to flat fielding, PSF variations, and ACS
pixel--to--pixel correlations in quadrature to the flux uncertainties
(Sirianni et al. 2005). Our average color uncertainties range from 0.01 to 0.03~mag. 

As an estimate of galaxy total magnitude, we used SExtractor's
MAGAUTO. This quantity is not a perfect estimator (see
discussion in Mei et al. 2009, and references therein), which must be borne in mind 
when comparing to other samples.

\section{Results}

\subsection{Morphology--density relation}

\begin{figure}
\centerline{\includegraphics[angle=0,scale=0.5]{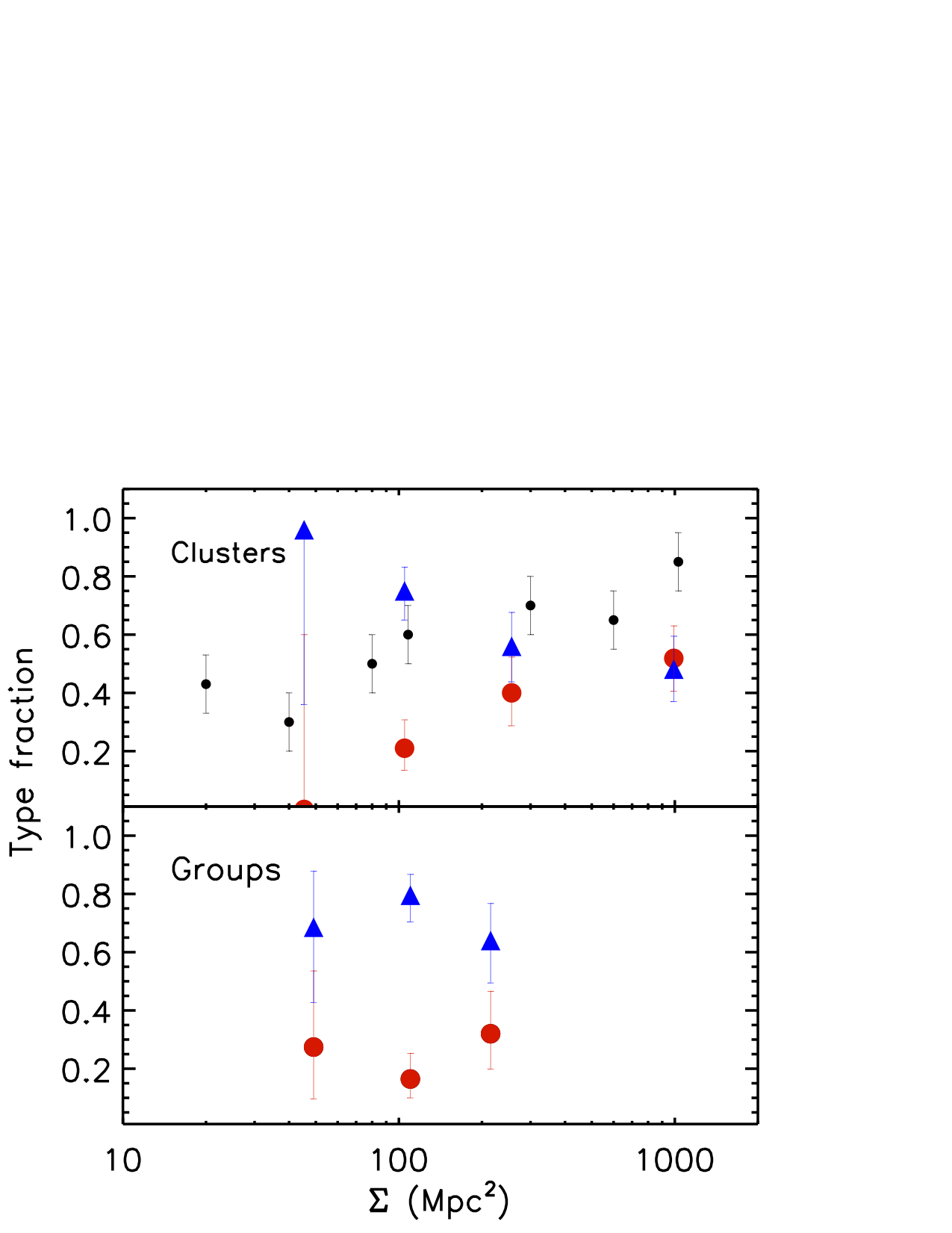}}
\caption{Morphology--density relation in the Lynx clusters (top) and
  groups (bottom) for our luminosity--limited sample. We use visual
  morphology to separate ETGs  from late--type galaxies. We show early--type (red circles) and late--type
  (blue triangles) fractions as a function of projected galaxy
  density (see text and notice that these symbols are different from
  those used in Fig.~\ref{fig4}). In the two clusters we observe an increase in early--type
  fraction with density, as expected in galaxy clusters,
  with a difference at the high density end. As a comparison, we show
 ETG fractions
  from the entire 7 cluster sample from the ACS Intermediate cluster
  survey (Postman et al. 2005) as small black dots. The entire sample shows an increase
  of the early-type population up to $\approx 80\%$. In the Lynx clusters, however, the early--type fraction does
  not exceed the late--type population fraction. In fact, they are
  found to contribute half of the entire population even at high
  densities. The group fractions are dominated by late--type galaxies,
  with fractions of early--types never exceeding $\approx 25\%$
  of the sample, a value consistent with field fractions and 
  Lynx cluster fractions at the same densities (see top figure). \label{fig6a}}
\end{figure}

\begin{figure}
\centerline{\includegraphics[angle=0,scale=.50]{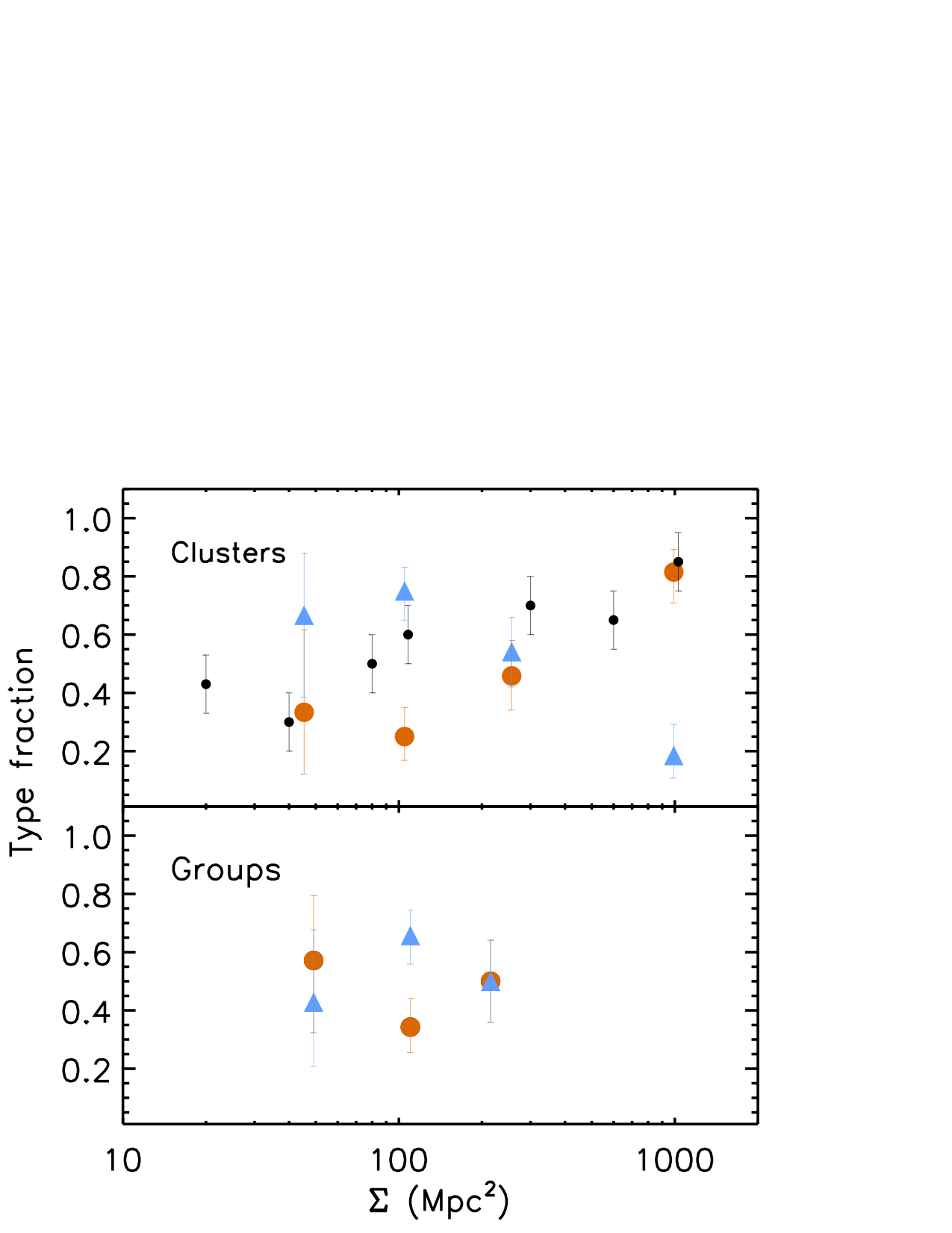}}
\caption{ Morphology--density relation for BDGs
  (orange circles) and late spirals (light blue triangles),  for our luminosity--limited sample.   In the
  two clusters, we see an increasing bulge
  fraction with density.  For comparison we show early--type fractions
  from the entire 7 cluster sample from the ACS Intermediate cluster
  survey (Postman et al. 2005) as small black dots. Compared to
  Fig.~\ref{fig6a}, the BDG population fraction in the clusters
  approaches that ETG fraction observed
  in the entire ACS Intermediate cluster survey from Postman et
  al. (2005). The group fractions show that  the BDG and late type spiral populations contribute each about
  half of the total population, and are closer to cluster
  fractions at the same densities (see top figure) and in agreement
  with group fractions at these redshifts. \label{fig6b}}
\end{figure}

Morphology--density relations in the Lynx clusters and
groups, as derived from our visual morphological
classification, are shown in Fig.~\ref{fig6a} and  Fig.~\ref{fig6b}. 
Here, we have counted early--type mergers as early--type,
and late--type mergers or unknown morphologies as late--type.
Projected densities were estimated following Postman et al. (2005). We
used a seven nearest neighbor algorithm with a $f_{corr}=3$. To
correct for background contamination, we subtract from galaxy type
counts an estimated number of non-cluster members, as $N^{c}_T = N_{total}f_{total,c}f_{c,T}$ (see Postman et
al. Appendix~B for details). $N_{total}$ is the total number of galaxies in a
density bin, $f_{total,c}$ the total number of contaminants, estimated
from our spectroscopic sample with the new photometric redshifts, and
$f_{c,T}$ as the contamination rate expected for each
morphological class.

 Fig.~\ref{fig6a}  shows the visual early (E, S0) and late--type fractions. In the two clusters, the early--type
  fraction increases with density, as expected in galaxy
  clusters. However the early--type fraction in the two Lynx clusters
  never increases beyond $\approx 50\%$, as compared to the $\approx
  80\%$ observed in massive clusters at $z\approx1$
  (Postman et al. 2005). In the top panel of Fig.~\ref{fig6a}, we show
  Postman et al. (2005) results for the entire seven cluster sample from
  the ACS IRCS. Compared to the entire sample, the
  early fraction in the Lynx clusters shows a further lack of
  early--type galaxies, with a confidence of $\sim 2 \sigma$. 
This is mainly due to a strong presence of late--type galaxies in the
Lynx~W sample. Note that Lynx~W was not used in the Postman
et al. (2005) analysis.

The group populations are dominated by late--type galaxies,
  with the fraction of early--types never exceeding $\sim 25\%$
  of the sample, a fraction consistent with that in the two Lynx
  clusters in regions of the same density as the groups, and similar to field fractions at these
  redshifts (e.g. Fig.~9 in Postman et al. 2005). Part of this is due
  to an overdensity of late--type galaxies in Group~1 (see
  Fig.~\ref{fig5}, top right of Group~1). We have verified that galaxy fractions don't
  change significantly even when this overdensity is excluded from 
  the analysis, with early--type fractions remaining close to $\sim 25\%$.

Since we cannot exclude that
diffuse features around Sa galaxies might be tidal structures or
merger remnants (instead of a disk), we estimated fractions of what we call {\it
  bulge--dominated  galaxies} (hereafter BDGs): visual early--type
galaxies, Es and S0s, plus BDS (BDS are the visual Sas by definition). In this way we want to verify if, when
added to the early--type galaxies, the galaxies
that we classify as Sas might be the galaxies we are missing when
counting only early--type galaxies.
In Fig.~\ref{fig6b},  we show the fraction
of BDGs and later--type (e.g. not counting the Sas) galaxies. When the
entire BDG population is shown, its fraction in the clusters increases
to that expected from the entire Postman et al. (2005) sample. In the
groups, BDG fractions contribute at $\approx 50\%$, more consistent
with ETG fractions observed in groups at these redshifts  (e.g. Poggianti et
al. 2009; Just et al. 2010). 

As observed in Mei et al. (2009), such a high fraction of BDS not observed in the
other clusters of the ACS IRCS. A similar
evolution of early--type spirals is found in the COSMOS field
(Bundy et al. 2010), where this population increases at $z \approx
1$ for galaxies with masses similar to those of our sample.
Since the fraction of early to late--type galaxies changes between
these two clusters and clusters at $z<1.2$, while the number of BDG
does not, as pointed out in Mei et al. (2009),  
these populations, whether they are spiral or early--type
galaxies with tidal features or merger remnants, could eventually
evolve into the ETG morphologies as we observe them at
$z\approx1$. We might speculate that a similar
evolution might have occurred in the groups, but with the small
statistics we have, we are unable to address this question.

\subsection{The color-magnitude relation in groups and clusters}

\begin{figure}
\centerline{\includegraphics[angle=90,scale=.60]{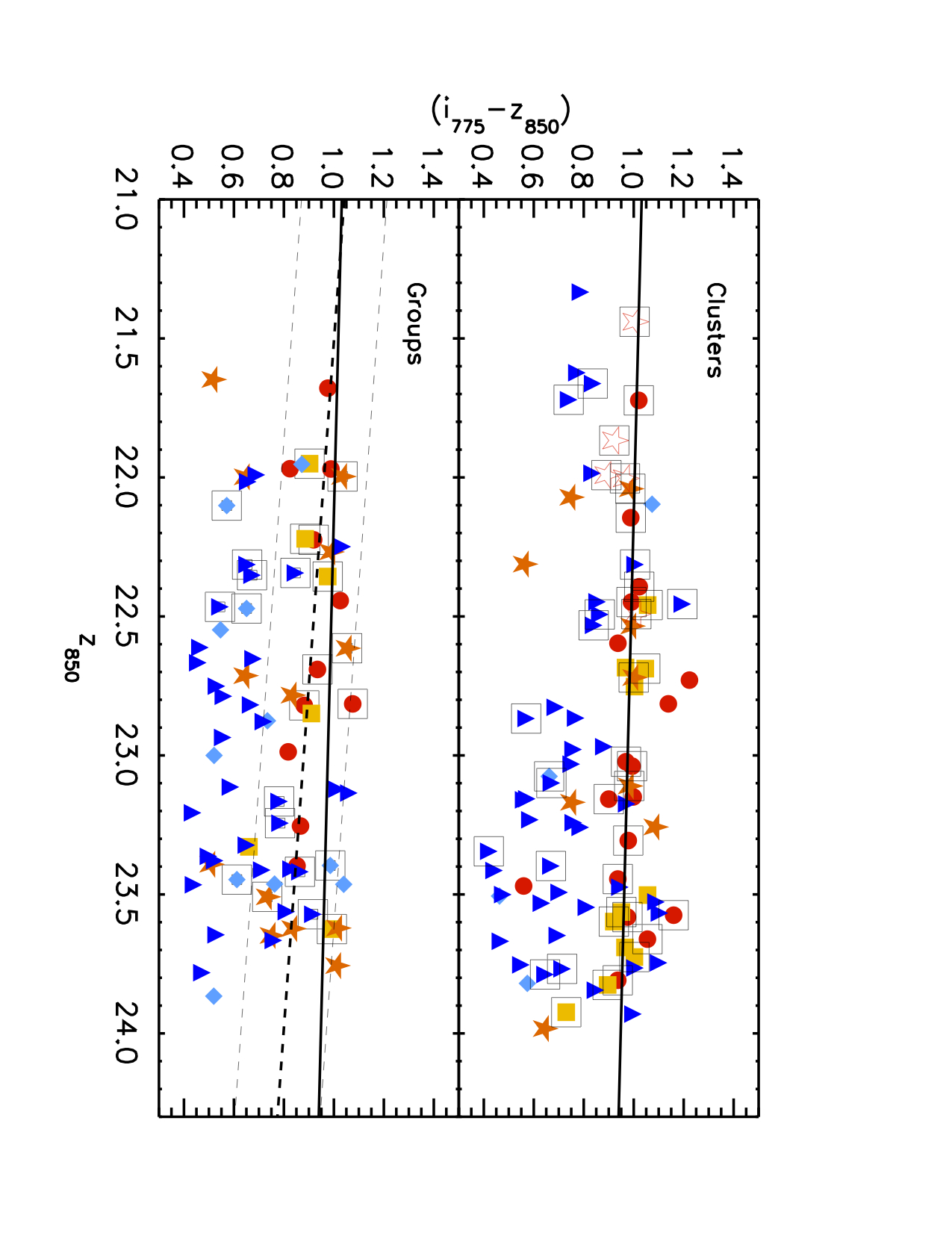}}
\caption{Color--magnitude relation for our cluster (top) and group
  (bottom) samples. Symbols are as in Fig.~\ref{fig4}. The continuous
  lines show the fit to the cluster elliptical red sequence, within one
  virial radius. The
  dashed lines in the bottom figure show the fit and the 3~$\sigma$
  scatter around the elliptical red sequence in the groups, within one
  virial radius. Double and triple mergers observed in the clusters
  all lie on the red sequence, and involve BDGs with luminosity
  larger than $L_*$. We do not observe BDG mergers in the
  groups. 4 disturbed objects are
  observed in the clusters, 9 in the groups. Only
  4 of these objects lie on the red sequence, and their luminosity
  is equally distributed between high and low luminosities.  \label{fig8}}
\end{figure}

The color--magnitude relations (CMRs) for the clusters and groups are shown in
Fig.~\ref{fig8}. Two of the bright spirals in the clusters are spectroscopically
confirmed, the brightest is in the very center of the Lynx~E. The fits to the red sequence have been performed as in Mei
et al. (2009) for different early--type populations in the clusters
and the groups. We fit three parameters:  the zero point, slope and
scatter around the red sequence:

\begin{equation}
(i_{775}-z_{850}) = Zero Point + Slope \times (z_{850} - 22.5)
\end{equation}

The CMR was fit using a robust linear fit based on Bisquare weights (Tukey's biweight; Press et al. 1992), and the uncertainties on the fit coefficients were obtained by bootstrapping on 1,000 simulations. The scatter around the fit was estimated from a biweight scale estimator (Beers, Flynn \& Gebhardt 1990) that is insensitive to outliers in the same set of bootstrap simulations.
A linear, least--squares fit with three-sigma clipping and standard RMS scatter gives similar results to the biweight scale estimator within $\approx$~0.001-0.002~mag for the slope and the scatter. 

To estimate the {\it intrinsic} galaxy scatter (i.e., not due to galaxy color measurement uncertainties), we estimated the additional scatter
needed beyond the measurement error to make the observed  $\chi^2$ per degree of freedom of the fit equal to one.
Again, the uncertainty on the internal scatter was calculated by bootstrapping on 1,000 simulations. 

Our results are given in Table~\ref{tab2} and  Table~\ref{tab2bis},
for all cluster and group selected galaxies and for the galaxies within one virial radius, respectively. The continuous
  lines in Fig.~\ref{fig8}  show the fit to the cluster elliptical red
  sequence within one virial radius. The
  dashed lines show the fit and the 3~$\sigma$
  scatter around the elliptical red sequence in the groups, within one
  virial radius. 
For the group ETG red sequence, we found lower zero points (at $\sim 2 \sigma$), and larger
scatters, both indicating a younger galaxy population 
 (e.g., Kodama \& Arimoto 1997; Kauffman \& Charlot 1998; Bernardi et al. 2005; Gallazzi et al. 2006).

To quantify these differences, we use predictions from stellar
population models (e.g. Mei et al. 2009). In Paper~II, we demonstrated that
galaxy age and mass predictions change significantly for galaxies with
ages around 1-2~Gyrs, when using stellar population models in which the contribution of the thermally pulsing asymptotic giant
branch (TP-AGB) phase is better modeled (Maraston 2005 - M05; Charlot \&
Bruzual  2007-CB07 ). Since Lynx galaxies  lie at an epoch in which the
Universe is $\approx$~5~Gyrs old, and we have shown the early-type
galaxies  to have 
formation epochs between 1-4 Gyrs (Paper~II and III), it is
important that we quantify differences in ages with both BC03 and
models with a better accounting of the TP-AGB, in order
to test the stability of our results.

Using simple BC03 and CB07 (similar to MA05) stellar population models, we find 
similar results.  As in
Mei et al. (2009), we consider three models: 1) a simple, {\it single burst} solar metallicity CB07 stellar population model; 2) a model with solar metallicity and {\it constant star formation rate} over a time interval $t_1$ to $t_2$, randomly chosen to lie between the age of the cluster and
the recombination epoch; 3)  a model with solar metallicity
and with an {\it exponentially decaying star formation rate}.
Models (1), (2) and (3) yield similar results, as already noted in Mei et
al. (2009).

With both the {\it single burst} and the {\it constant star formation rate}
model, we find the average luminosity-weighted age for
ellipticals and ETGs in the
groups to be  $\sim 0.5 - 0.7$~Gyr younger than the cluster ellipticals, similar to
peripheral ellipticals, and S0 galaxies in the two clusters (Mei et
al. 2009).   A model with solar metallicity
and an {\it exponentially decaying star formation rate}, however, predicts that if
galaxies with larger scatters had a different star formation history
(exponential decay versus single burst), they would have an average
luminosity-weighted age similar to the cluster ellipticals. This
scenario also predicts bluer color residuals, as observed in our
groups. 

Galaxies in the groups have then either formed in a short episode of
star formation (approximated here as a single burst) later than
galaxies in the clusters, or formed at the same time, but following a
star formation history with longer timescales. 
Fitting galaxy spectral energy
distributions in Paper~II, we studied galaxy ages and masses in detail,
and showed that discriminating between these two
hypotheses depends on the adopted stellar
population model (see Fig. 9 of that paper). 
Using M05 and CB07, we find older populations in
the two clusters with respect to the groups and the field, in
agreement with results from the CMR.
In Paper~III,  we show how, using BC03 models, we obtain different exponential
decaying times in the clusters and the field.

A difference between the age in groups and clusters, if it
exists, is small when compared to the differences in age in galaxies of
different masses in the same sample, as pointed out and discussed in Paper~II. 

The double and triple bulge mergers observed in the clusters
  all lie on the red sequence, and involve galaxies with luminosity
  larger than the characteristic $L_*$ ($m_* \approx 23$~mag at
  z$\approx 1.3$; Mei et al. 2006b). We do not observe red bulge mergers in the
  groups. Moreover, 4 disturbed objects are
  observed in the clusters, and  9  in the groups. Only
  4  (1 in the clusters, 3 in the groups) of these objects lie on the
  red sequence (they show clumps, irregular structure and are probably
  reddened by dust), and their luminosity
  is equally distributed to high and low luminosities. 

Luminous red bulge mergers occur in clusters,
while both the clusters and the groups show disturbed objects that indicate late--type
interactions at all luminosities, and mainly in the blue cloud.
These disturbed objects show late--type merging, tidal features and
complex morphology that suggest a higher incidence of galaxy
transformations in the 
groups, in terms of mergers and other galaxy
interactions, mainly happening in the blue cloud. The 4 disturbed objects in the
cluster are not all in peripheral regions, but lie mostly at
around half of the virial radius.

BDGs have mostly red colors, e.g. they tend to lie on the red
sequence. While their red colors might indicate a lack of star formation,
it could be also hidden by the presence of dust. 
We estimated their star formation rate using galaxy spectral energy
distribution (SED) fitting, following 
our procedure as described in Paper~II. 
On the red sequence, their star formation rates are always estimated at less than $\approx
25 M_{\sun}/year$ and $18 M_{\sun}/year$, in the clusters and the
groups, respectively. One BDG in Lynx~W  has a star formation rate of $66
M_{\sun}/year$ and lies at $>3 \sigma$ (bluer) than the best fit to
the ETG red sequence. In the groups, two BDGs have star formation rates $>
100 M_{\sun}/year$, and both lie at $>3 \sigma$ (bluer) than the best fit to
the ETG red sequence, and one BDG has star formation rate of $45
M_{\sun}/year$ and lies at $>2.3 \sigma$  (bluer) than the best fit to
the ETG red sequence.  BDGs are therefore mostly passive.

\begin{figure}
\includegraphics[angle=90,scale=.35]{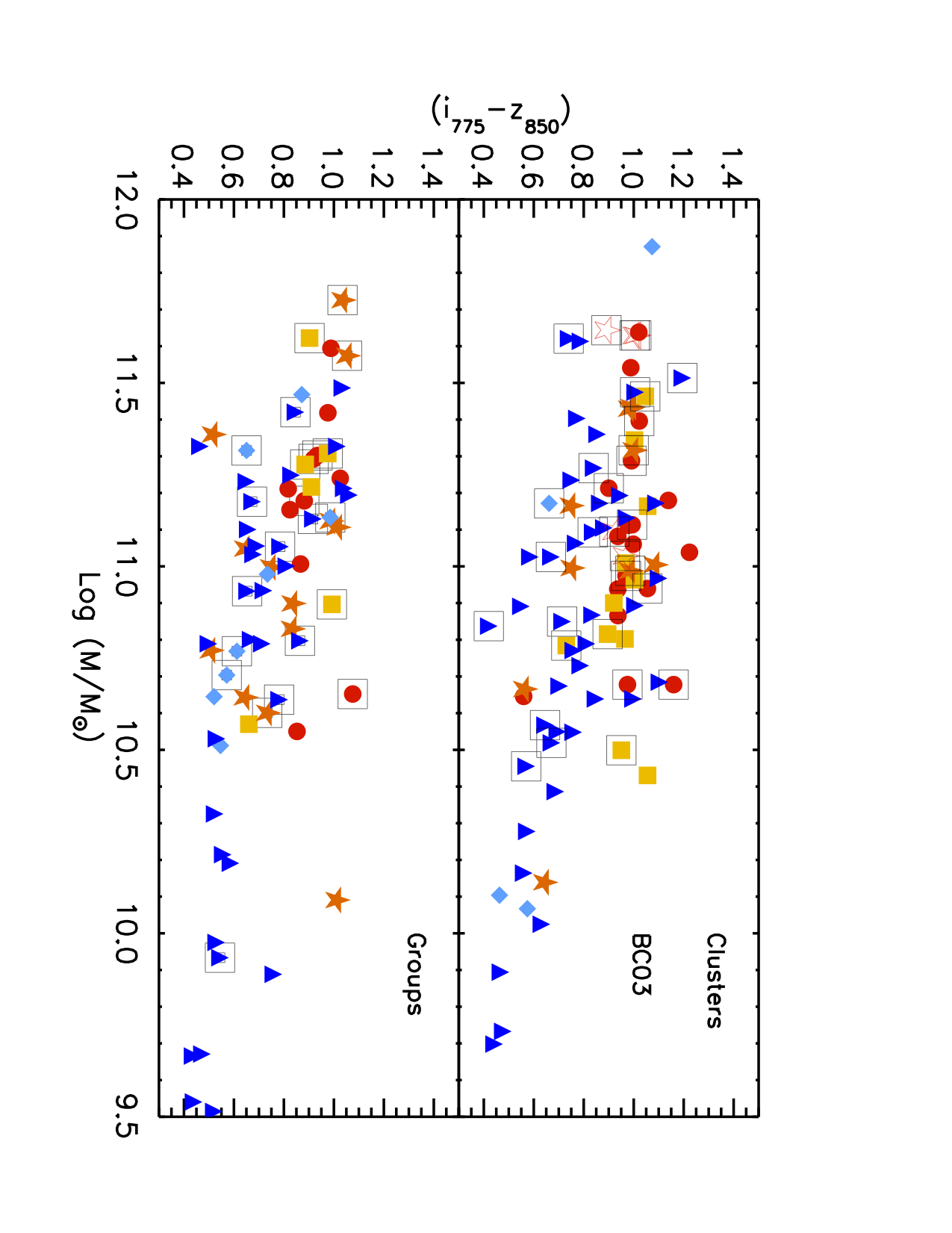}
\includegraphics[angle=90,scale=.35]{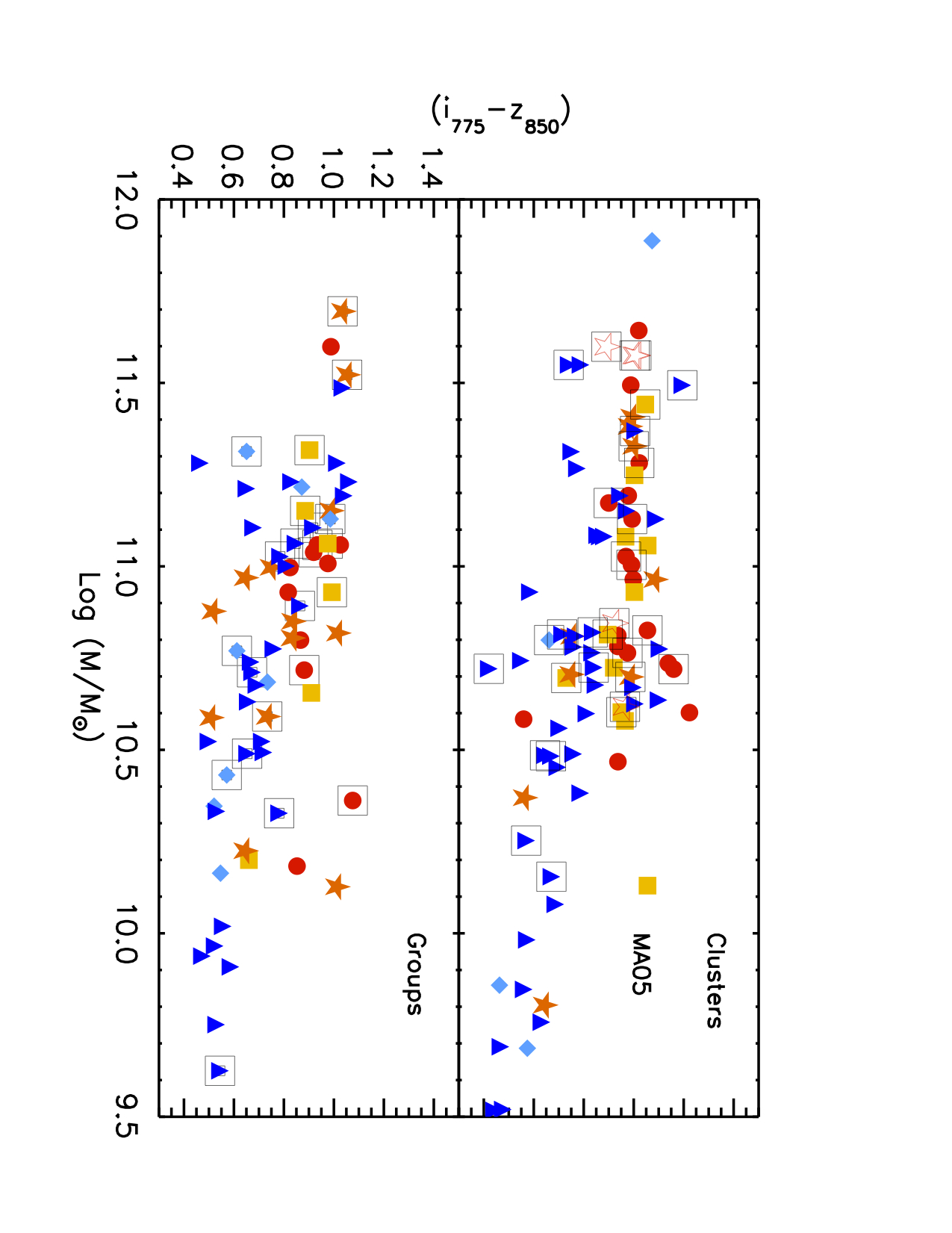}
\includegraphics[angle=90,scale=.35]{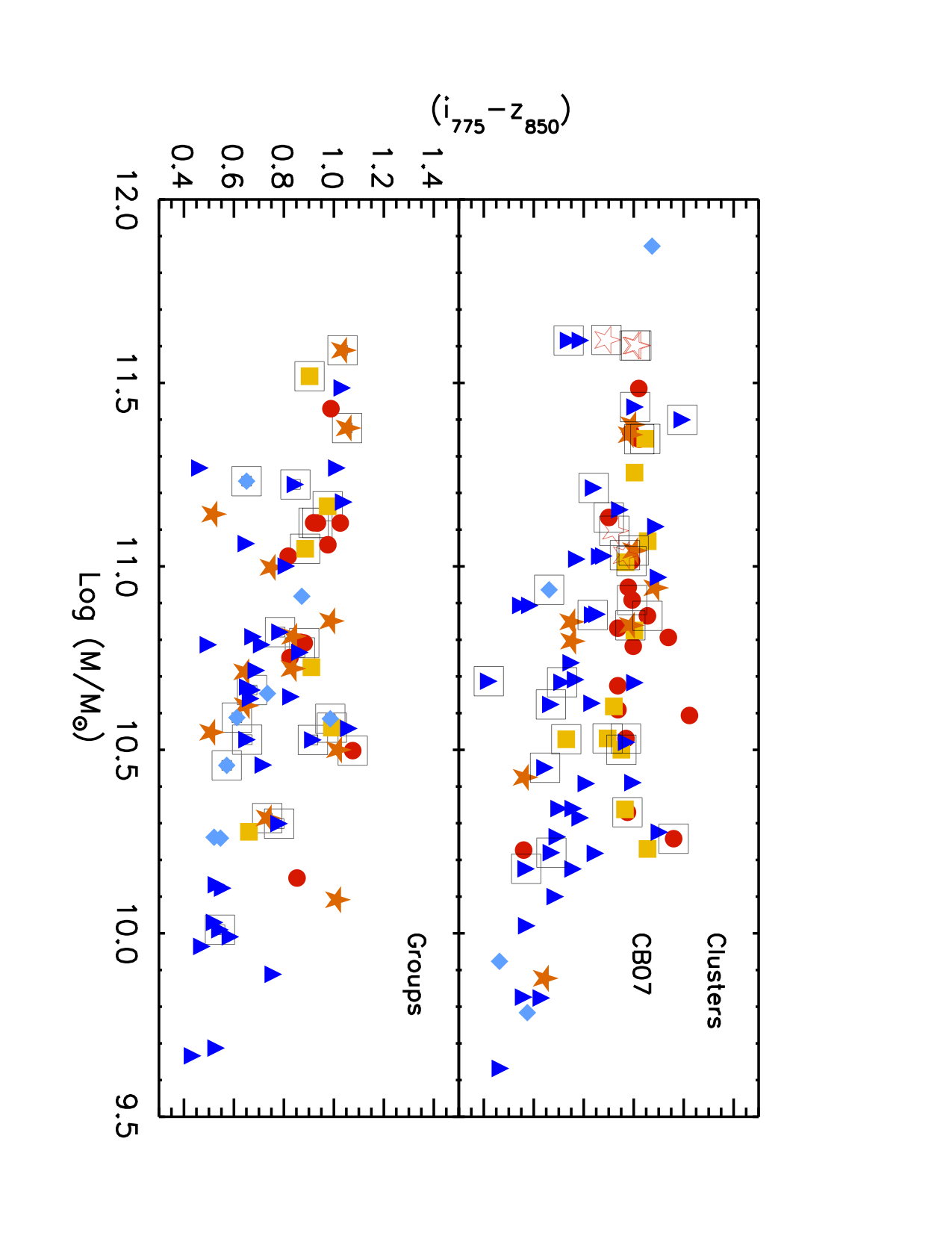}
\caption{Color--stellar mass relation for our cluster (top of each panel) and group
  (bottom of each panel) samples. Symbols are as in Fig.~\ref{fig4}. Masses are
  estimated by SED fitting using templates from three stellar
  population models (BC03, MA05 and CB07; see text for details). \label{fig9}}
\end{figure}

In Fig.~\ref{fig9}, we plot colors as a function of stellar
mass, and can define a mass--limited sample. Galaxy stellar masses were
derived by galaxy SED fitting, as described in
Paper~II, using templates from the three different stellar
population models: BC03, M05, CB07.

As expected, galaxies with the same luminosity range in the
color--magnitude diagram have different masses, since the
mass--to--light ratio differs between quiescent and star forming
galaxies. As discussed in Paper~II, ranges in mass are different because we are using different stellar
population models: the BC03 model overestimates masses and ages to
account for the missing light from the TP-AGB phase (see Paper~II for
a detailed analysis at our redshift, and MA05).

If we define mass--limited samples, some of the spiral galaxies that
were previously included in the luminosity limited sample are now
excluded. If we use as a conservative lower mass limit in our
early--type sample $M=10^{10.6} M_{\sun}$, $\approx$~30\% (44\%/57\%) of the
spirals have lower masses when using BC03 (MA05/CB07). At a mass limit
of $M=10^{10.4} M_{\sun}$, $\approx$~20\% (30\%/39\%) of the
spirals have lower masses when using BC03 (MA05/CB07).

\subsection{Morphological content of the red sequence}

\begin{figure}
\begin {tabular} {r@{\hspace*{-3mm}}r}
\includegraphics[angle=0,scale=.40]{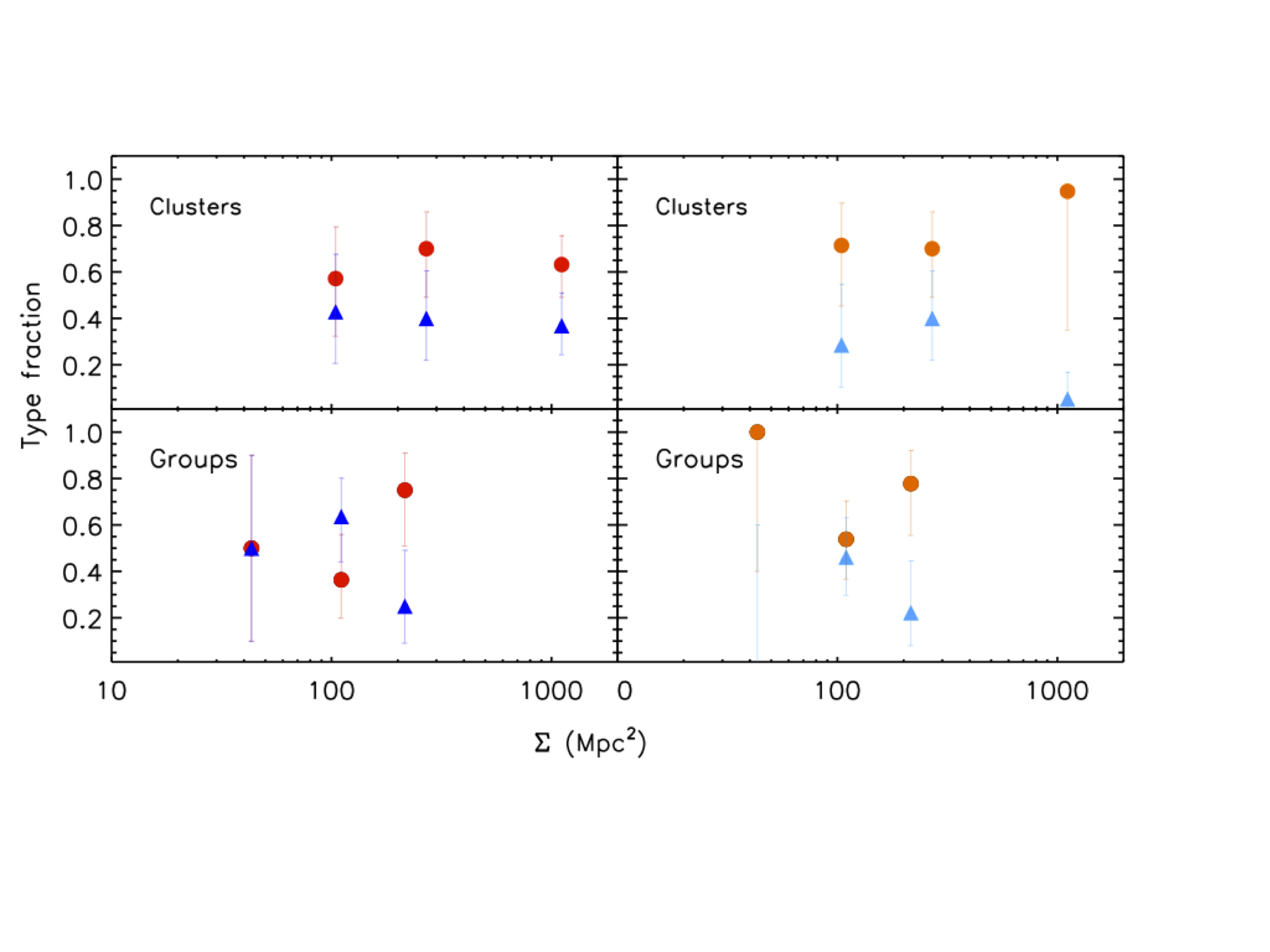} 
\end{tabular}
\caption{Morphology--density relation on the red sequence. We use the
  same symbols as in Fig.~\ref{fig6a} and Fig.~\ref{fig6b}.
On the left, the fraction of early (red circles) and late--type galaxies
(blue triangles) on the red sequence is
around 0.5 in groups and low density cluster regions, in agreement
with the results for the entire sample in  Fig.~\ref{fig6a}. The
percentage of ETGs grows in intermediate density regions. In
high density regions in clusters, though, we again find a fraction of about
a half. This is due to the presence of Sa galaxies in these high density
regions, three of which are involved in the triple merger in Lynx~W. On
the right, we show the fraction of BDGs (orange circles) and late
spirals (light blue triangles). The BDG fraction in clusters on the
red sequence shows the same kind of percentages observed in the overall
sample, with BDGs dominating the high density end. For the groups, the relative
fractions of BDG and late spirals are similar to the early to
late--type fractions. 
\label{fig10}}
\end{figure}

The morphological content of the red sequence is shown in Fig.~\ref{fig10}. On the left, we compare early
to late--type fractions in the clusters and groups. On the right, we show the BDG to
later--type spiral fraction on the red sequence. As pointed out when analyzing total fractions, ETG fractions
on the red sequence are low in the two Lynx clusters when compared
to average ETG fractions observed in the other
ACS Intermediate Cluster survey clusters at $z<1.2$ (see also Mei et
al. 2009). The red sequence
in the groups is dominated by late--type galaxies.

Comparing the fraction of BDG and later--type spirals, the BDG
fraction is close to the fraction of
ETG observed in the other clusters at $z<1.2$, and
the groups show results consistent with the ETG fractions in
clusters at the same projected density and group fractions at
$z \approx 1$ (George et al., 2011). The BDS population in both clusters and groups seems to
compensate for the lack of an early--type population. This suggests that
at $z\approx 1$, they might evolve into the early--type population on
the red sequence, or that they will leave the red
sequence later in time.

\subsection{Morphological fractions in a mass limited sample}

\begin{figure}
\includegraphics[angle=0,scale=.40]{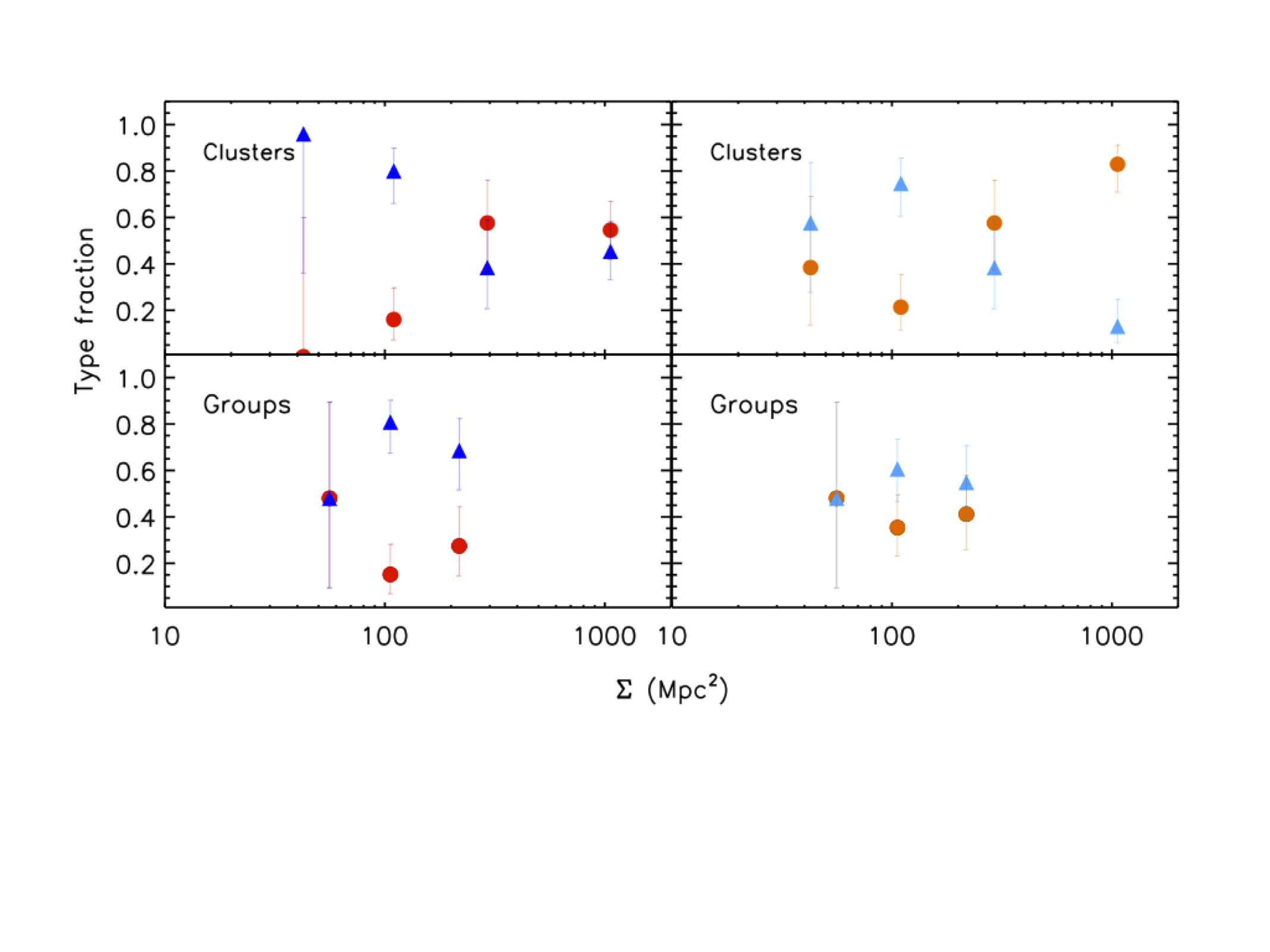}
\caption{Morphology--density relation in a mass
limited sample, for galaxies with mass  $M>10^{10.6}
M_{\sun}$. We use the
  same symbols as in Fig.~\ref{fig6a} and Fig.~\ref{fig6b}). On the left, we show the ETG (red circles) and late--type (blue triangles) fractions. On
the right, we show the BDG (orange circles) and 
 late spiral
(light blue triangles) fractions. Masses are
  estimated by SED fitting using templates from CB07. Our results do
  not change significantly using the other two stellar population
  models. \label{fig12}}
\end{figure}

Fig.~\ref{fig12} shows early and late-galaxy fractions in a mass--limited sample. We selected galaxies with $M>10^{10.6} M_{\sun}$, to
be conservative with respect to the completeness of our ETG
sample.
The uncertainties on morphological fractions are calculated following
Gehrels (1986; see Section 3 for binomial statistics; see also Mei et
al. 2009). 
These approximations apply even when ratios of
 different events are calculated from small numbers, 
and yield the lower and upper limits of a binomial distribution 
within the 84\% confidence limit, corresponding to 1$\sigma$.

In the mass--limited sample, we observe the same trends as in the
luminosity-limited sample: the fraction of ETGs is
lower than that observed in the rest of the ACS IRCS. BDG fractions,
though, show once again fractions that are similar
to those of ETGs in the ACS IRCS.
Considering galaxies with $M>10^{10.6} M_{\sun}$, $R<R_{200}$, and projected density $\Sigma$ larger than a given $\Sigma_{lim}$, we obtain the total
early-type/BDG fractions as given in Table~\ref{tab3}.

Holden et al. (2007) show that in mass--limited
samples, early-type fractions can change dramatically with respect to
those in luminosity--selected samples (e.g. while in both elliptical
fractions do not show significant evolution, in luminosity selected
samples S0s show evolution). In fact, their work demonstrates that
the early-type fraction of galaxies with $M>10^{10.6} M_{\sun}$ does
not evolve from z=0.83 to the present (opposite to the evolution
shown by the luminosity--selected cluster samples from e.g. Postman et al. 2005,
Desai et al. 2007, Poggianti et al. 2009). At z=0.83, they report a fraction
of $89 \pm 7 \%$ for ETGs in regions of density $\Sigma
> 500 Mpc^{-2}$, at that mass limit.

\begin{figure}
\centerline{\includegraphics[angle=90,scale=.50]{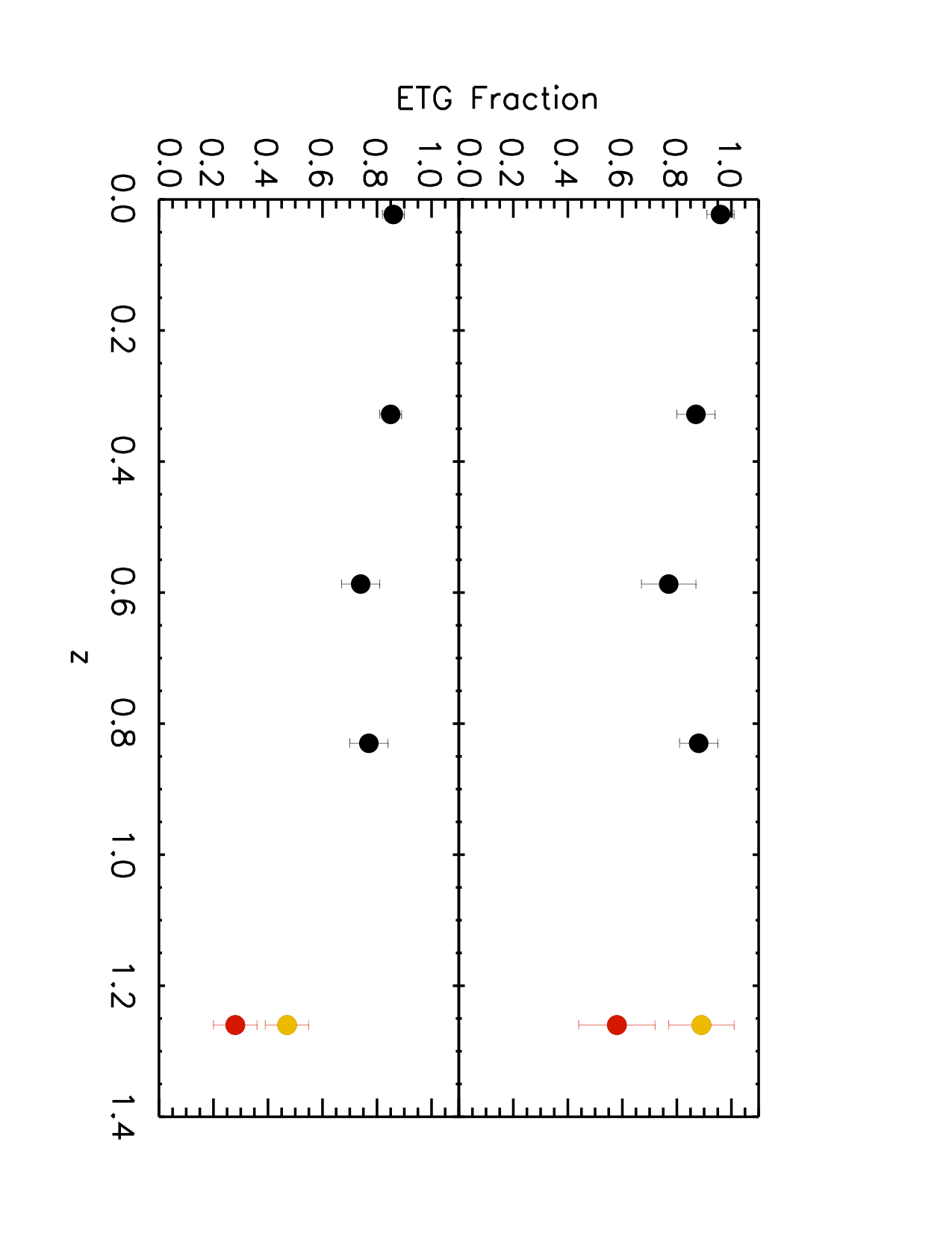}}
\caption{Evolution of the ETG fractions from $z\sim
  1.3 $ to the present for the clusters (top), and the groups as
  compared to the clusters (bottom). The results from this work (BC03) are compared to those
  from Holden et al. (2007) for a sample of clusters at $z<1$ (black circles), with galaxies selected in the same way in
  mass ($M>10^{10.6} M_{\sun}$ ) and
 in the same density region ($\Sigma
> 500 Mpc^{-2}$). Lynx ETG and BDG fractions are shown by red and
yellow circles, respectively. Cluster ETG fractions do not show a
significant
evolution up to $z\sim 1.3$. When adding BDS, the results do not
change. Group fractions (as expected) are lower
than those in the clusters. \label{evolution}}
\end{figure}

Fig.~\ref{evolution}, compares our results to Holden et
al. (2007). In the Lynx clusters, we do observe a lack of ETGs in
our mass--limited sample, selected in a manner similar to Holden et
al. (2007), with an early-type fraction of $\approx 60 \pm 10 \%$ independent
of the stellar population model used for mass estimation.
However, when estimating the BDG fraction, we obtain  $\approx 89 \pm
10 \%$, consistent with the ETG fraction found from Holden et
al. (2007) at z=0.83. 

These results are consistent with a lack of a significant evolution in the fractions of
ETGs with masses $M>10^{10.6} M_{\sun}$.  We obtain similar results for a mass
limit of $M>10^{10.4} M_{\sun}$. We expect our photometric redshift sample to be
contaminated at $\sim 20\%$. Even in the extreme
case in which all early--types are real cluster members (i.e., no 
contamination), and $\sim 20\%$ of the spirals are not, we would
obtain fractions of $ \approx 65 \pm 10 \%$, and  $\approx 95 \pm
10 \%$, for the cluster ETGs and bulges, respectively.

\subsection{Galaxy sizes for different morphological types}

The analysis of evolution in morphological fractions has led us to
the hypothesis that at z$\approx$1.3, we observe a BDS population
that, losing their disk (or merger remnants or tidal features if these
galaxies are not spirals but merger remnants or ETG with tidal features),
might have evolved into an ETG population, to obtain the ETG
fraction observed in clusters at $z<1.2$. 

In this scenario, we should also observe a similarity in galaxy
size between BDS and ETG galaxies.
As we discuss in depth in Paper~IV, while field ETGs at $z\approx 1$, selected from the GOODS-CDFS,
show a distribution in sizes similar to the local cluster distribution (notice this might be different than the field local distribution, we will discuss further this point in future work), Lynx
cluster ETGs are more compact than local galaxies at a given mass (refer to Paper~IV for the quantitative analysis). These results, consistent with recent results from Valentinuzzi et
al. (2010), lead us to conclude that, while we do not need an evolution
in the overall mass--size relation for field ETGs, cluster ETGs require an evolution of their
mass--size distribution from $z \approx 1$ to the present. As mechanisms of
evolution, we propose that either part of the cluster ETGs become
larger with time and/or arise  from transformations of cluster late--type population
(non--ETG progenitors), and/or a new population of larger ETGs is
accreted onto the cluster.

\begin{figure}
\centerline{\includegraphics[angle=90,scale=0.6]{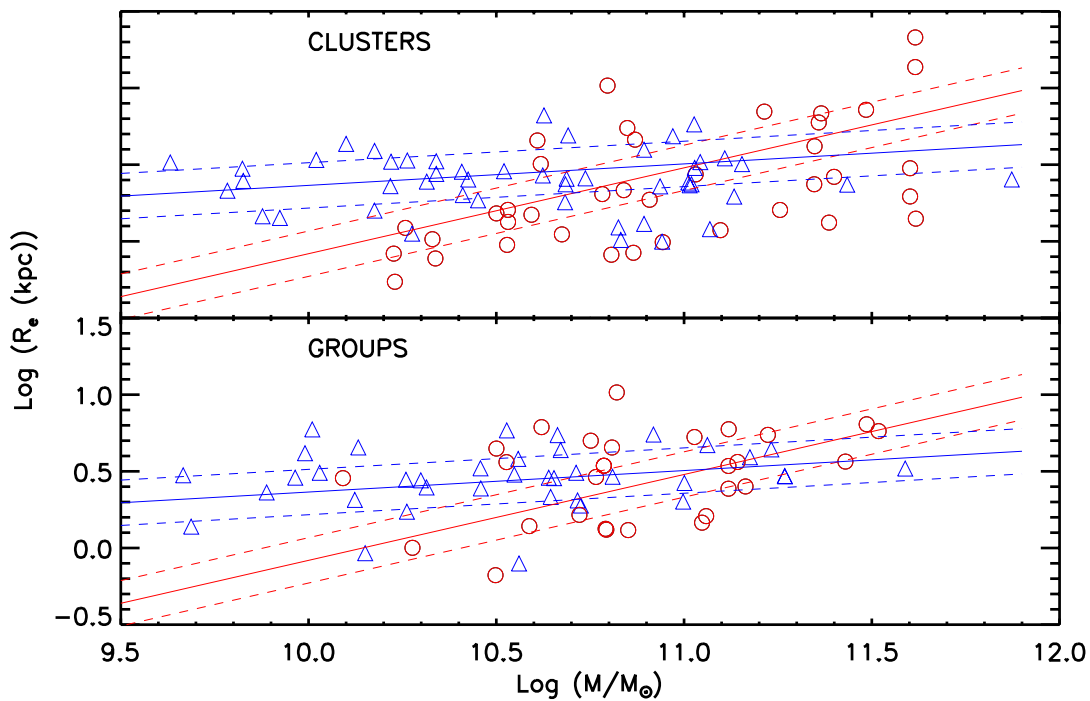}}
\caption  {The mass-size relation for all galaxies in our
sample (cluster and group galaxies on the top and the bottom,
respectively), compared to the local relation from the SDSS. The red (blue) continuous and dashed lines are the Shen et
al. (2003) ETG (disk) local relations and their 1~$\sigma$ dispersion,
respectively. Morphological types in our Lynx sample are selected as in Shen et
al. (2003) (notice that this is not the visual morphological
classification we have user so far):
red empty circles are galaxies with  Sersic index $n>2.5$ and empty
blue triangles are galaxies with $n<2.5$
Using the same morphological class definitions based on Sersic
index shows that both cluster and group galaxy mass--size relations do
not evolve.  
 \label{fig13}}
\end{figure}

In this paper, we explore late--type sizes to understand if part of
the larger ETG local population might have evolved from BDS.
In Fig.~\ref{fig13}, we plot the mass--size relation for our cluster and
group sample,  compared to the standard SDSS local relation from Shen et
al. (2003). In this
figure (and hereafter) we use stellar masses derived with CB07.  Our results are the same when using masses derived with BC03
and MA05. Shen et al. (2003) use as morphological classification a
simple separation in Sersic index $n$ or compactness. We plot their local
relation for early ($n>2.5$) and late  ($n<2.5$) galaxies and compare
to our results using exactly the same classification (see
Fig.~\ref{fig4} and Fig.~\ref{sersic}) (notice that here we are not
using our visual morphological classification as in all other analysis
in this paper). We find that
neither galaxies in the clusters nor in the groups show evolution
in the overall mass--size relation. The mass--size relation in the Lynx
superstructure is similar to the Shen et al. (2003) mass--size
relation.

\begin{figure}
\centerline{\includegraphics[angle=90,scale=.6]{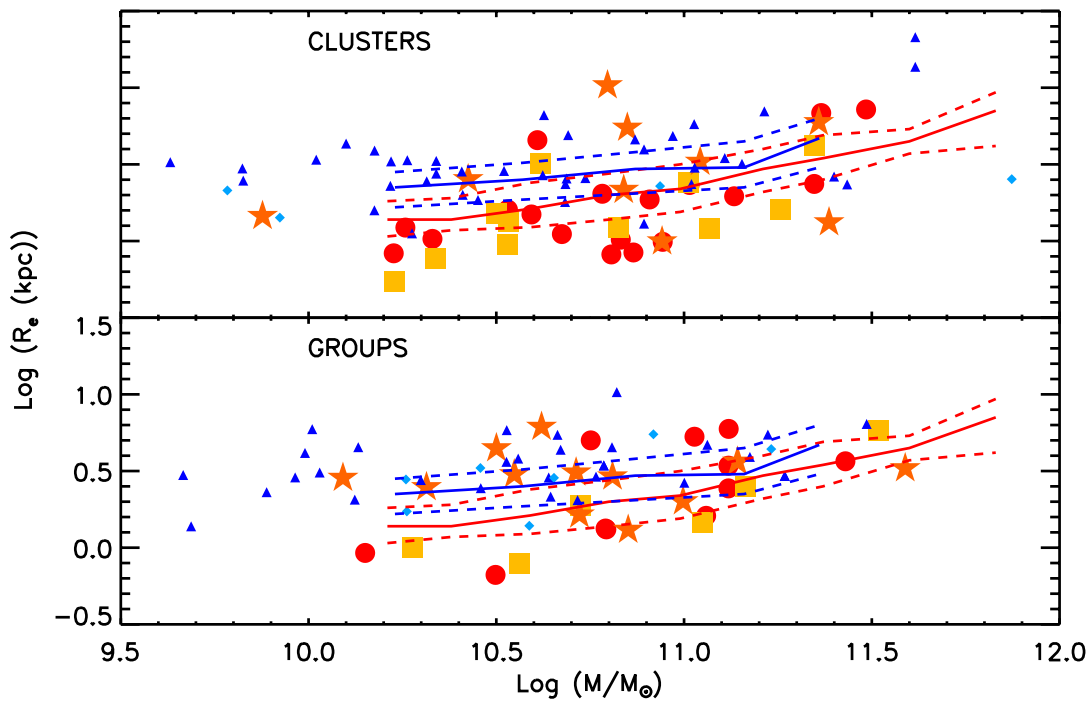}}
\caption  {The mass-size relation for all galaxies in our
sample (cluster and group galaxies on the top and the bottom,
respectively). Our visual morphological types are shown with the same symbols as in
Fig.~\ref{fig4}. The red (blue) continuous and dashed lines are the
Valentinuzzi et al. (2010) ETG (disk) local relations for WINGS clusters and their 1~$\sigma$ dispersion,
respectively. Valentinuzzi et al. morphologies were based on a visual 
classification, similar to our classification of the Lynx
galaxies. Cluster ETGs show a size distribution on average
smaller than the  Valentinuzzi et al.  ETGs (Raichoor et al. 2012). Lynx Sa galaxies (orange stars), for which the fit to a Sersic profile is
dominated by their large bulges, show a size distribution larger  than
the Lynx ETGs. The Lynx late-type galaxies have a distribution
similar to the Valentinuzzi et al. disk distribution. \label{fig15}}
\end{figure}

However, a simple morphological classification based on one structural
parameter is highly contaminated by a mixing of different morphologies,
as shown in Section~4,
and does not correspond to a visual
morphological classification. 
 In Fig.~\ref{fig15},
we therefore compare galaxies separated into our visual morphological
classes (as used in the other analysis in this paper). As a local sample, we use the morphological classified sample
from Valentinuzzi et al. (2010), corrected for masses estimated with a Salpeter IMF.

When a visual morphological classification is used at both high and
low redshift, ETGs in both
clusters and groups show a size distribution more compact than
ETGs in the local sample (see Paper~IV for a detailed analysis of the ETG Kormendy and mass--size relations in our
sample), while spirals do not show evolution in their overall
mass--size relation (when their
sizes are estimated by a single Sersic fit). 

While our Lynx galaxies do lie within 3~$\sigma$ from the average
Valentinuzzi et al. (2010) mass--size relation, at a given mass, the two gaussian
distributions are statistically
different for masses $M < 2 \times 10^{11} M_{\sun}$ which dominate our sample. Using results from
Table~1 of Paper~IV, and a Kolmogorov-Smirnov
 and Kuiper statistical test
for galaxies with masses $M < 2 \times 10^{11} M_{\sun}$, from all three
stellar population models, we obtain a
probability close to zero that the
Valentinuzzi et al. (2010) Gaussian distribution and ours are 
driven by the same distribution.
In the analysis of the ETGs in Paper~IV,
we concluded that to obtain the local ETG mass--size distribution, the
ETG population
in clusters and groups must have either accreted larger ETGs, or formed
new ones from non-ETG progenitors or have
gone through transformations that enlarged their own size, e.g. by minor
dry mergers (Naab et al. 2009, Shankar et al. 2011), or stellar
winds and/or quasar feedback (Fan et al. 2008).

Most of our Sas do show larger sizes than ETGs, and their size
distribution is in agreement with the hypothesis
that (at least part of) this population would transform into
early--types (even if this does not demonstrate that they will). This
new population might enlarge
the ETG mass--size distribution and reproduce that observed in the local
sample. Larger samples are needed to understand the relative
importance of this possible transformation with respect to the accretion of
larger galaxies and galaxy--galaxy transformations
(major/minor--dry/wet mergers and tidal interactions).

\section{Discussion and conclusions}

We have examined galaxy color and morphology in the
clusters and the groups of the
Lynx supercluster at $z \sim 1.3$. Our primary aim is to identify
galaxy properties and their transformation in the cluster and group
environment at redshifts where typical clusters are still being assembled. 

We confirm the detection of three groups in the Lynx supercluster (Nakata et al. 2005), by XMM Newton
X-ray imaging  and spectroscopic
follow--up at Subaru, Keck and Gemini telescopes.
From XMM Newton X--ray imaging, we detect Group~3 at 2.6~$\sigma$ and Group~2 at
1.6~$\sigma$. Group~1 shows a 0.5~$\sigma$ excess, and there is
 2~$\sigma$ excess of the X-ray emission close to it. From the X-ray
emission and scaling relations (Markevitch et al. 1998) we obtain a mass estimate for Groups~2 and 3 of $5.7 \pm
2.6  \times 10^{13} M_{\sun}$
and $5.0 \pm 1.7 \times 10^{13} M_{\sun}$, respectively, and a 95\% upper limit for the
mass of Group~1 at $M < 4.5 \times 10^{13} M_{\sun}$. 
From our spectroscopic runs in the groups, we obtained redshifts for the three confirmed groups:
$z=1.262 \pm 0.007$ (Group~1; from 9 members), $z=1.260 \pm 0.006$
(Group~2; from 7 members), $z=1.263 \pm 0.005$ (Group~3; from 9
members).

 These measurements confirm that these three groups are at the same
redshifts as Lynx~E and Lynx~W, the central clusters of the Lynx supercluster (see the Introduction). Group~1 is very close to
Lynx~W, and their galaxies are connected by a FoF algorithm. We
consider the group as a separate structure, however, because its center
lies at $\sim R_{200}$ from the center of the
cluster and extends to $\sim 2 R_{200}$, with an area of very low density between $0.5$ and 1~ $
R_{200}$. The group might be close to merging, or in the
merger process. Our present XMM imaging does not permit us to detail  
this interaction.

We classify galaxies in the groups as early--type (E, S0; ETG),
late--type (Sa, Sb and Sc), irregular and
disturbed (possible mergers, tidal features), using the same visual
classification as used for the two main clusters by Postman et
al. (2005). We compare our visual classification to automated methods. 
We show that if we would have used the log(A) vs log(C) plane or the
Sersic index to separate early from late--type galaxies, our ETG
sample would have been contaminated (especially by compact bulge--dominated
spirals, Sas) at the 30\% and 40\% level, respectively (see also Capak
et al. (2007)). More sophisticated automated classification (Huertas--Company et al. 2008, 2009, 2011) show only a 5\% contamination.

From the analysis of cluster and group galaxy morphologies and colors, we find that
while the color--magnitude relation is already in place, this
supercluster shows high fractions of red, bulge--dominated spirals (BDS). We
classify BDS as Sa
galaxies,
e.g., spirals with large bulges but with a
clearly visible diffuse component that we identify as a disk but that
might also be due to merger remnants and tidal features.

From analysis of the cluster and group color--magnitude relation,
we found lower zero points (e.g., bluer average colors), and larger intrinsic scatters in the
ETG CMR for the groups, both indicating a younger galaxy population or more complex star formation
histories with respect to the cluster ellipticals (e.g. Mei et
al. 2006ab, 2009). 
We quantified this difference in age/star formation history with simple stellar population models,
and found the average luminosity-weighted age of ellipticals
in the groups to be $\sim 0.5-0.7$~Gyr younger than cluster
ellipticals. This difference in age is similar to that between the
cluster core population and the peripheral ellipticals and S0s in the two clusters
(Mei et al. 2009). 

Our ETG sizes, ages and star formation histories are analyzed in 
detail in three companion papers.
In Paper~II, we study cluster and group ETG formation epochs
as a function of stellar mass and compare them to the field. In Paper~III, we
compare ETG ages and star formation histories in the two main clusters and
the field. Paper~IV  examines the ETG
Kormendy and mass--size relations. 
In Paper~II and Paper~III, we study galaxy
stellar populations in detail. We find that the dependence of age on
mass (older galaxies are the most massive) is more important than the
dependence of age on environment (Paper~II). 
Star formation histories in the clusters are less
extended than in the field (Paper~III). We also show (Paper~II) the
importance of understanding limitations of stellar population models when
interpreting results. 

We find
that the early--type
  fraction increases with the density in the two clusters, as in other galaxy
  clusters at this redshift (e.g. Postman et al. 2005). However, the ETG fraction in the two Lynx clusters
  never rises above $\approx 50\%$, as compared to the $\approx
  80\%$ observed in the other massive clusters at $z\approx1$
  (Postman et al. 2005; Smith et al. 2005; Desai et al. 2007). In the
  groups, ETG fractions never exceed $\approx 25\%$, 
  consistent with cluster fractions at the same densities and  field
  fractions at the same redshift (Postman et al. 2005),  and lower than
  what has been observed in groups 
  (Poggianti et al. 2009; Just et al. 2010)

This lack of ETGs is compensated by the large number of red early
spirals, the Sas. When we measure the bulge fractions (E+S0+Sa), we
obtain fractions similar to those observed for ETGs in clusters at
$z\sim1$. Within $0.6 \times R_{200}$ (the same region used by Poggianti et al. 2010), our cluster and group ETG fractions are $56 \pm 9$\% and
$23 \pm 9$\%, respectively, while BDG fractions are $70 \pm 9$\% and
$37 \pm 9$\%, respectively. 

When selecting our sample by mass, our results do not change. 
For galaxies with masses $M>10^{10.6} M_{\sun}$ within  $\Sigma
> 500 Mpc^{-2}$, we find  an
ETG fraction of $\approx 60 \pm 10 \%$ and $\approx 30 \pm 10
\%$ in the clusters and groups,
respectively, and independent
of the stellar population model used for the mass estimation.
However, when we estimate the BDG fraction in the two clusters we obtain  $\approx 90 \pm
10 \%$. 

We compare these results to the Holden et al. (2007) ETG
fractions at z=0.83 and lower redshifts, where the galaxy selection was
made in the same way in mass and by density regions. Our results show that the ETG
and bulge fractions are still high at $z\sim1.3$. We do not
see a significant evolution of the ETG and the BDG fraction either overall, or on the
red sequence, from $z\sim 1.3 $ to the present for galaxies with masses  $M>10^{10.6} M_{\sun}$ within  $\Sigma
> 500 Mpc^{-2}$. Our results do not change if we use  $M>10^{10.4}
M_{\sun}$ as the mass limit.

As observed in Mei et al. (2009), a high fraction of BDS is not observed in the other clusters in the ACS
IRCS. Since the fraction of ETG galaxies is changing between these two clusters and
clusters at $z < 1.2$, but the BDG fraction does not, as we pointed out in Mei et al. (2009),
this BDS population might be thought to eventually evolve
into ETGs.
Since BDS  are also red and passive, they must have
had, firstly, their star formation quenched and then subsequently experience a morphological transformation.

Recent results based on the COSMOS field (Bundy et al. 2010) show a
similar trend: an increase of massive ($M \gtrsim  10^{11}M_{\sun}$)
ETGs and a decline of $M \lesssim 10^{11}M_{\sun}$ BDSs (in this
case selected by their sersic index $1.25<n<2.5$) on the red sequence,
from
$z\approx$~1 to present.  These authors deduced that at
least $60\%$ of the BDSs were transformed into
ETGs (e.g. S0s) on the scale of 1-3~Gyrs, and that these
transformations might occur as a single major merger event or through
multiple evolutionary stages, including disk disruption by minor
mergers or accretion of cold gas in star--forming galaxies. These scales are similar
to the time scales of a few Gyrs derived from measurements of the S0
velocity dispersions in clusters
(e.g. Moran et al. 2007). 

The transformation of spirals into S0s has been largely discussed in the
literature (Bekki et a. 2002; Christlein et al. 2004; Postman et al. 2005; Poggianti et al. 2009,2010; Wilman et
al. 2009; Bekki \&
Couch 2011). The most studied 
idea is that as they are accreted into
galaxy clusters, spirals lose their disks and become early--type galaxies. This
hypothesis is difficult to reconcile with the bulge luminosity
 of local cluster late--type galaxies (Christlein et al. 2004).

The BDS that we observe, though, have luminous bulges and show an interesting
mass--size relation.
To test the hypothesis that ETGs might have been formed from these
progenitors,  we 
estimated the mass--size relation for our entire sample and compared
it to the local relation from the SDSS. 
We do so by comparing, on the one hand, samples selected by Sersic index,
and on the other hand, 
samples selected based on visual morphology. Our results 
differ in the two cases. If we would have used a simple
separation of early from late--type, identifying ETGs as galaxies
with a Sersic index $n>2.5$, the two classes would
have lied on the local mass--size relations of galaxies selected
in the same way in the SDSS. Samples of galaxies selected on their
Sersic index do not show evolution in their mass--size relation.

A simple morphological classification based on one structural
parameter is, however, highly contaminated by a mixing of different
morphologies, with compact spirals (mainly Sas) being identified as
ETGs (see Section~4). In the second case, we compare our sample to the SDSS sample, using in both
cases a visual (or in case of the SDSS, trained on visual)
morphological classification. As we discuss in depth in Paper~IV, we find that Lynx
cluster ETGs are more compact than local galaxies at a given
mass (see also different results from Cooper et al. 2012 and Papovich
et al. 2012). We show in this paper that spirals, however, lie on the local mass--size relation.
We conclude in Paper~IV that to obtain the local mass--size distribution, galaxies
in clusters and groups must have either accreted larger ETGs or have
gone through transformations that enlarged their own size (e.g. Fan et
al. 2008; Naab et al. 2009; Shankar et al. 2011), or have formed ETGs from later type
progenitors.

Most of our Sa galaxies do show larger sizes than ETGs, and their size
distribution is in agreement with the hypothesis
that if (at least part of) this population transformed into
early--types, this would enlarge
the ETG size distribution and reproduce that observed in the local
sample. Larger samples are needed to understand the relative
importance of this transformation with respect to the accretion of
larger ETGs from the field,  the role of dry/wet minor mergers, wet
major mergers or strangulation of
later type progenitors.


Since we do not have dynamical data for those Sa galaxies, we cannot
exclude the hypothesis that the diffuse component we observe and
visually identify as a disk, might be
tidal structures or merger remnants that will fade with time. A detailed dynamical analysis of
the Sa sample would help in understanding the disk nature of the
diffuse component that we observe.


Compared to field samples, Paper~IV have shown that field ETGs at
 $z\sim1.3$ and with similar mass limits to the present sample, show a
 mass--size relation in agreement with the local relation. Cassata et
 al. (2011) have shown that, in the field, ETGs enlarge their size and
 increase their stellar mass by a factor of 5 between $z=2$ and 1. At
 z$\sim1.3$, field galaxies are already on the local mass--size
 relation, while cluster ETGs are about twice as compact  on average, indicating that their size distribution will increase 
 (see also Strazzullo et al. 2011 for a similar results at z=1.4, but
 also Papovich et al. 2012 for a different point of view, perhaps
 because our sample selections are different).

At the redshift of the Lynx supercluster, $\Lambda CDM$ models predict that environmental
effects should not yet be predominant (e.g. McGee et al. 2009). We are
 observing clusters before they accrete a significant
 number of field and group galaxies and groups before they infall into
 clusters. Our cluster population should  be a pristine population at
 $z\sim1.3$,  before a significant
fraction of group and field galaxies is accreted onto clusters.

Our results show a cluster population with high percentages of
BDS or bulges with a significant diffuse
component (tidal structures/merger remnants; see above) that
could evolve into ETGs. When we consider the fraction of this
population together with the current cluster ETG population, we obtain
BDG fractions that are in agreement with local BDG fractions. If
these objects do transform into ETGs (e.g. S0 galaxies), the pristine
ETG cluster population will
not need to accrete and transform additional spirals to reach an ETG
population consistent with the local population. From the observed double and
triple red merger, some (we observe a few) of those red ETGs
will merge to build more massive ETGs. Our results then suggest that, at these redshifts, the spiral galaxies that transform into S0s (see e.g. Postman et al. 2005, Desai et al. 2007, Poggianti et al. 2010) are mainly Sas.

The groups will eventually be accreted onto the the main clusters to
form a single more massive structure (with a mass typical of clusters in the local Universe). Our groups show
pre--processing in terms of wet similar--size mergers, the presence of
small companions in two galaxies, and tidal tails and clumpy morphologies
in $\sim 10$\% of the sample. Group BDG fractions are around
$40-50$\%. If we assume pre--processing in groups in terms of
transformation of bulge--dominated spirals into ETGs, we would need
part of the disturbed morphologies to be also transformed into ETGs for
the group accretions to not significantly change the total ETG fractions in
clusters, as observed at lower redshift. From our sample, it appears
that simple transformations of BDGs into
ETGs would be in agreement with the mild evolution found for galaxies  with masses  $M>10^{10.6} M_{\sun}$ within  $\Sigma
> 500 Mpc^{-2}$ up to $z\sim 1.3$. Further morphological
transformations that would significantly change (e.g. at $\sim20$\%)
the ETG fraction
should happen at lower masses and/or at lower densities.  

From the analysis of the group red sequence, group ETGs also show
red sequence zero point and scatter that are consistent with those of peripheral ellipticals and S0
galaxies in clusters at $z\approx1$ (Mei et al. 2009).

\section{Summary}

Our work aimed to confirm the detection of groups in the Lynx supercluster, and study galaxy morphology and stellar population at a time when typical clusters are being assembled.

Our main results are the following:

\begin{itemize}
\item With XMM Newton imaging and spectroscopic follow-up at Subaru, Keck and Gemini telescopes, we confirm the detection of three groups around the two Lynx clusters, RX~J0849+4452 and
RX~J0848+4453 at redshift
$z=1.262 \pm 0.007$ (Group~1), $z=1.260 \pm 0.006$
(Group~2), and $z=1.263 \pm 0.005$ (Group~3). The estimated group masses are $M < 4.5 \times 10^{13} M_{\sun}$ (Group~1), $5.7 \pm
2.6  \times 10^{13} M_{\sun}$ (Group~2),
and $5.0 \pm 1.7 \times 10^{13} M_{\sun}$ (Group~3).

\item We morphologically classify galaxies by visual inspection. Our early–type galaxy (ETG) sample would have been contaminated at the 30\% –40\% level by simple automated classification methods. We observe wet mergers in both clusters and groups, and dry mergers only in clusters.

\item The Lynx supercluster shows low fractions of ETGs (when compared to other clusters at similar redshift). This lack of ETGs is compensated by high fractions of red, bulge--dominated spirals (BDS).  These objects have large bulges and a diffuse disk component that shows features similar to arms. These features might also be due merger remnants and/or tidal features. Since the fraction of ETG galaxies changes between the Lynx clusters and
clusters at $z < 1.2$, but the BDG fraction does not  (Mei et al. 2009), this might be evidence that
the BDS population will eventually evolve
into ETGs (see also Bundy et al. 2010). BDS  are also red and passive: if these transformations will happen at later time, BDS must have
had, firstly, their star formation quenched and then subsequently experience a morphological transformation. Some of these BDS are observed in dry mergers.  If
these objects do transform into ETGs (e.g. S0 galaxies), the pristine
ETG cluster population will
not need to accrete and transform additional spirals to reach an ETG
population consistent with the local population.  Our results then suggest that, at these redshifts, the spiral galaxies that transform into S0s (see e.g. Postman et al. 2005, Desai et al. 2007, Poggianti et al. 2010) are mainly Sas.

\item In the Lynx clusters, the ETG fraction of galaxies with masses  $M>10^{10.6} M_{\sun}$ within  $\Sigma
> 500 Mpc^{-2}$ do not show evolution from $z\sim 1.3 $ to the present. Our results do not change if we use  $M>10^{10.4}
M_{\sun}$ as the mass limit. This confirms the results to the Holden et al. (2007) for the evolution of ETG
fractions up to z=0.83

\item We studied the cluster and group color--magnitude relation (CMR). The CMR is already in place in both environments. We find lower zero points (e.g., bluer average colors), and larger intrinsic scatters in the ETG CMR for the groups.
Using simple stellar population models, we find that this difference corresponds to 
average luminosity-weighted ages of ellipticals
in the groups to be $\sim 0.5-0.7$~Gyr younger than cluster
ellipticals. This result is consistent with the analysis of the galaxy SEDs in Raichoor et al. (2011) and Rettura et al. (2011).
However, the dependence of age on
mass is more important than the
dependence of age on environment (Raichoor et al. 2011).

\item We study the mass--size relation for galaxy samples selected using their Sersic index and visual morphology. When using the Sersic index, our results do not show an evolution of the galaxy mass-size relation, while when using visual morphology, at a given mass the ETG sizes are smaller than local cluster galaxies (see also Raichoor et al. 2012). Our result shows that galaxy samples that are morphologically selected as early based on Sersic index are highly contaminated by a mixing of different visual
morphologies and can lead to biases in results on the mass--size relation evolution. The spiral galaxy  mass--size relation do not show significant evolution. At a given mass, BDS show a large range of sizes, that span both the spiral and ETG sizes. Most BDS show large sizes though, and their size
distribution is in agreement with the hypothesis
that if (at least part of) this population transformed into
ETGs, it would enlarge
the ETG size distribution and reproduce the ETG mass--size relation observed in the local
sample.

\end{itemize}

\acknowledgments

ACS was developed under NASA contract NAS 5-32865. This research 
has been supported by the NASA HST grant GO-10574.01-A, and Spitzer
program 20694.
The {Space Telescope Science
Institute} is operated by AURA Inc., under NASA contract NAS5-26555.
Some of the data presented herein were obtained at the W.M. Keck
Observatory, which is operated as a scientific partnership among the
California Institute of Technology, the University of California and
the National Aeronautics and Space Administration. The Observatory was
made possible by the generous financial support of the W.M. Keck
Foundation.  The authors wish to recognize and acknowledge the very
significant cultural role and reverence that the summit of Mauna Kea
has always had within the indigenous Hawaiian community.  We are most
fortunate to have the opportunity to conduct observations from this
mountain. Some data were based on observations obtained at the Gemini Observatory, which is operated by the
Association of Universities for Research in Astronomy, Inc., under a cooperative agreement
with the NSF on behalf of the Gemini partnership: the National Science Foundation (United
States), the Science and Technology Facilities Council (United Kingdom), the
National Research Council (Canada), CONICYT (Chile), the Australian Research Council
(Australia), Ministério da Ci\^encia e Tecnologia (Brazil) 
and Ministerio de Ciencia, Tecnolog\`ia e Innovaci\`on Productiva
(Argentina), Gemini Science Program ID: GN-2006A-Q-78. RD gratefully acknowledges the support provided by the BASAL Center for Astrophysics and Associated Technologies (CATA), and by
FONDECYT grant N. 1100540. We thank the anonymous referee
for the very constructive suggestions and Shannon Patel for useful discussions.




{\it Facilities:}  \facility{HST (ACS), Spitzer (IRAC, MIPS), KPNO:2.1m
(FLAMINGOS), Hale (COSMIC), Keck:I (LRIS), GEMINI(GMOS)}

\clearpage




\clearpage



\clearpage

\begin{table*}
\begin{center}
\caption{Lynx supercluster  \label{tab1}}
\vspace{0.25cm}
\resizebox{!}{2.cm}{
\begin{tabular}{llccccccccccccccc}
\tableline \tableline\\
Cluster/Group & z&$\sigma_v^{a}$ &$L^{X_a}_{bol}$&$R_{200}^b$
&$M_{tot}^{X_c}$ &$N_{zphot}^{d}$&$N_{zspec} ^{d}$\\
&&(km/s)&($10^{44} $  erg/sec) &Mpc&$10^{14} M_{\odot}$&&\\ \hline
RX~J0849+4452&1.261&$740^{+113}_{-134}$ &$2.83\pm0.17$&0.9&$2.9\pm1.5$&49&19\\
RX~J0848+4453& 1.270& $650 \pm 170$&$1.04
\pm0.73$&0.8&$1.4\pm1.0$&40&22\\
Group~1&1.262 &&$<0.20 \pm 0.09$&$<0.5$&$< 0.45 \pm 0.14$&31&9\\
Group~2&1.260&&$0.28 \pm 0.22$&0.5&$0.57 \pm
0.26$&15&7\\
Group~3&1.263&&$0.23 \pm 0.13$&0.5&$0.50 \pm
0.17$&28&9\\
\tableline \tableline\\
\end{tabular}}
\end{center}
\small{
$a$:  Velocity dispersion for the clusters come from Jee et al. (2006) and Stanford et
al. (2001). We do not estimate velocity dispersions for the
groups. If the groups were virialized, using Calberg et al. (1997), we
would obtain characteristic velocity dispersions of
$\approx$~400~km/s. Bolometric luminosities were derived within an
over-density $\Delta_z=500$ for an Einstein--de Sitter universe from
Ettori et al. (2004). \\
$b$: $R_{200}$ refers to the radius at which the cluster mean density is 200 times the critical density and is derived from the cluster velocity dispersion (Carlberg et al. 1997). \\
$c$: Total masses were estimated out to $R_{500}$ using a cluster $\beta$ model together with the measured emission--weighted X--ray  temperature (Ettori et al. 2004)\\  
$d$ : $N_{zphot}$ and $N_{zspec} $ are the photometrically and
spectroscopically-selected members for each structure, respectively
}
\end{table*}

\begin{table*}
\begin{center}
\caption{CMR fit results for all our selected galaxies.  \label{tab2}}
\vspace{0.25cm}
\resizebox{!}{4cm}{
\begin{tabular}{llccccccccccccccccc}
\tableline \tableline\\
$Sample$ &$Type$&$N^{gal}$&$Zero \  point$&$Slope$&$Scatter$ \\
&&&mag&&mag\\
    Clusters & E+S0 & 26 &       1.01  $\pm$       0.01 &     -0.027  $\pm$    0.019 &    0.058  $\pm$    0.012  \\ 
                           & E & 16 &       0.99  $\pm$       0.01 &     -0.014  $\pm$    0.028 &    0.065  $\pm$    0.020  \\ 
                          & S0 & 10 &       1.03  $\pm$       0.02 &     -0.060  $\pm$    0.031 &    0.036  $\pm$    0.013  \\ 
                 Lynx E & E+S0 & 19 &       1.00  $\pm$       0.02 &      0.001  $\pm$    0.032 &    0.063  $\pm$    0.014  \\ 
                            &E & 12 &       0.99  $\pm$       0.02 &      0.023  $\pm$    0.033 &    0.064  $\pm$    0.018  \\ 
                          & S0 &  7 &       1.00  $\pm$       0.03 &     -0.020  $\pm$    0.041 &    0.037  $\pm$    0.016  \\ 
                  Lynx W &E+S0 &  9 &       1.00  $\pm$       0.02 &     -0.056  $\pm$    0.018 &    0.028  $\pm$    0.015  \\ 
                           & E &  6 &       0.98  $\pm$       0.02 &     -0.046  $\pm$    0.034 &    0.025  $\pm$    0.023  \\ 
                          & S0 &  3 &       1.05  $\pm$       0.01 &     -0.139  $\pm$    0.054 &    0.011  $\pm$    0.021  \\ 
                  Groups& E+S0 & 15 &       0.92  $\pm$       0.02 &     -0.019  $\pm$    0.050 &    0.078  $\pm$    0.010  \\ 
                           & E & 11 &       0.93  $\pm$       0.04 &     -0.096  $\pm$    0.072 &    0.074  $\pm$    0.019  \\ 
                          & S0 &  4 &       0.95  $\pm$       0.01 &      0.073  $\pm$    0.013 &    0.029  $\pm$    0.026  \\ 
\tableline \tableline\\
\end{tabular}}
\end{center}
\end{table*}

\begin{table*}
\begin{center}
\caption{CMR fit results within one virial radius. \label{tab2bis}}
\vspace{0.25cm}
\resizebox{!}{4cm}{
\begin{tabular}{llccccccccccccccccc}
\tableline \tableline\\
$Sample$ &$Type$&$N^{gal}$&$Zero \ point$&$Slope$&$Scatter$ \\
&&&mag&&mag\\
     Clusters & E+S0 & 23 &       1.00  $\pm$       0.01 &     -0.032
     $\pm$    0.016 &    0.037  $\pm$    0.008 \\
                           & E & 14 &       0.99  $\pm$       0.01 &
                           -0.028  $\pm$    0.014 &    0.025  $\pm$
                           0.010 \\
                          & S0 &  9 &       1.03  $\pm$       0.02 &
                          -0.063  $\pm$    0.040 &    0.039  $\pm$
                          0.015 \\
                 Lynx E & E+S0 & 16 &       0.99  $\pm$       0.02 &
                 -0.001  $\pm$    0.028 &    0.036  $\pm$    0.008 \\
                            &E & 10 &       0.97  $\pm$       0.02 &
                            0.012  $\pm$    0.025 &    0.033  $\pm$
                            0.010 \\
                          & S0 &  6 &       1.00  $\pm$       0.04 &
                          -0.014  $\pm$    0.072 &    0.045  $\pm$
                          0.024 \\
                  Lynx W &E+S0 &  9 &       1.00  $\pm$       0.02 &     -0.057  $\pm$    0.020 &    0.027  $\pm$    0.015\\
                           & E &  6 &       0.98  $\pm$       0.02 &     -0.044  $\pm$    0.036 &    0.025  $\pm$    0.023\\
                          & S0 &  3 &       1.05  $\pm$       0.01&     -0.140  $\pm$    0.054 &    0.011  $\pm$    0.021\\
                  Groups& E+S0 &  8 &       0.92  $\pm$       0.03 &      0.016  $\pm$    0.063 &    0.066  $\pm$    0.024\\
                           & E &  5 &       0.92  $\pm$       0.06 &     -0.081  $\pm$    0.106 &    0.057  $\pm$    0.030\\
                          & S0 &  3 &       0.95  $\pm$       0.01 &      0.066  $\pm$    0.002 &    0.011  $\pm$    0.021\\

\tableline \tableline\\
\end{tabular}}
\end{center}
\end{table*}

\begin{table*}
\begin{center}
\caption{Early-type and BDG fraction for a mass--limited sample with
  $M>10^{10.6} M_{\sun}$, $R<R_{200}$, and $\Sigma <
  \Sigma_{lim}$. \label{tab3} }
\vspace{0.25cm}
\resizebox{!}{2.5cm}{
\begin{tabular}{llccccccccccccccccccccccc}
\tableline \tableline\\
$Sample$ &$Type$&$\Sigma_{lim}$&$N^{gal}_{BC03}$&Frac. BC03 &$N^{gal}_{MA05}$&  Frac. MA05  &$N^{gal}_{CB07}$&
Frac. CB07\\
&&& $Mpc^{-2}$&$\%$&$\%$&$\%$\\
     Clusters & E+S0  &80 &  26&     43  $\pm$       7 & 25&46  $\pm$   0.08&20& 43 $\pm$    8\\
& E+S0 &500 &   11&    58  $\pm$       14 & 11&58 $\pm$ 14 &10&56  $\pm$ 14\\
                           & E +S0 +Sa&80 &     35&  57  $\pm$       7   &33&60$\pm$ 7 &28&60  $\pm$  8\\
                         & E+S0 +Sa& 500 & 17&      89  $\pm$       12 &  17&                   $89^{+7}_{-13}$& 16& $89^{+7}_{-13}$\\
                  Groups & E+S0 & 80 & 13&      28  $\pm$       8  & 12&32      $\pm$   9 &11&  22  $\pm$       9 \\
                           & E +S0 +Sa& 80 &   22&      47  $\pm$       8 &18&49  $\pm$       9 &17&35  $\pm$       9\\
\tableline \tableline\\
\end{tabular} 
}
\end{center}
\end{table*}


\clearpage
\begin{thebibliography}{}
\bibitem[Abraham, Valdes, Yee, \& van den  Bergh(1994)]{ab94}      
Abraham, R.~G., Valdes, F., Yee, H.~K.~C., \& van den Bergh, S.\ 1994, \apj, 432, 75 

\bibitem[Arnouts et al.(2002)]{a02} 
Arnouts, S., et al.\ 2002, \mnras, 329, 355 


\bibitem[Beers, Flynn, \& Gebhardt(1990)]{b90} 
Beers, T.~C., Flynn, K., \& Gebhardt, K.\ 1990, \aj, 100, 32 

\bibitem[Bekki et al.(2002)]{be02} 
Bekki, K., Couch, W.~J., \& Shioya, Y.\ 2002, \apj, 577, 651 


\bibitem[Bekki \& Couch(2011)]{be11} 
Bekki, K., \& Couch, W.~J.\ 2011, \mnras, 415, 1783 



\bibitem[Bernardi et al. (2005)]{ben05}
Bernardi, M., Sheth, R. K., Nichol, R. C., Schneider, D. P., Brinkmann, J. 2005, \mnras, 129, 61


\bibitem[Bernardi et al.(2010)]{be10} 
Bernardi, M., Shankar, F., Hyde, J.~B., Mei, S., Marulli, F., \& Sheth, R.~K.\ 2010, \mnras, 404, 2087 



\bibitem[Bertin, E.~\& Arnouts 1996]{ba96}
Bertin, E.~\& Arnouts, S.\ 1996, \aaps, 117, 393 

\bibitem[Bielby et al.(2010)]{bi10} 
Bielby, R.~M., et al.\ 2010, \aap, 523, A66 


\bibitem [Blakeslee et al. (2003)]{bla03}
Blakeslee, J.P., Franx, M., Postman, M. {\it et al.} 2003a, ApJL, 596, 143 

\bibitem[Blakeslee et al.(2003)]{blab03} 
Blakeslee, J.~P., Anderson, K.~R., Meurer, G.~R., Ben{\'{\i}}tez, N., \& Magee, D.\ 2003b, ASP 
Conf.~Ser.~295: ADASS XII, 257

\bibitem[Boselli \& Gavazzi(2006)]{bo06} 
Boselli, A., \& Gavazzi, G.\ 2006, \pasp, 118, 517 


\bibitem[Bruzual A.~\& Charlot(2003)]{BC2003}
Bruzual A., G.~\& Charlot, S.\ 2003, \mnras, 344, 1000 (BC03)


\bibitem[Bundy et al.(2010)]{bu10} Bundy, K., et al.\ 2010, 
\apj, 719, 1969 



\bibitem[Calzetti et al.(2000)]{ca00} Calzetti, D., Armus, 
L., Bohlin, R.~C., Kinney, A.~L., Koornneef, J., 
\& Storchi-Bergmann, T.\ 2000, \apj, 533, 682 



\bibitem[Capak et al.(2007)]{ca07} Capak, P., Abraham, 
R.~G., Ellis, R.~S., Mobasher, B., Scoville, N., Sheth, K., 
\& Koekemoer, A.\ 2007, \apjs, 172, 284 

\bibitem[Careberg (1997)]{cal97} Calberg et al. 1997, \apj, 478, 462

\bibitem[Cassata et al.(2011)]{ca11} Cassata, P., et al.\ 
2011, \apj, 743, 96


\bibitem[Conselice et al.(2000)]{co00} 
Conselice, C.~J., Bershady, M.~A., \& Jangren, A.\ 2000, \apj, 529, 886 

\bibitem[Cooper et al.(2012)]{2012MNRAS.419.3018C} Cooper, M.~C., Griffith, 
R.~L., Newman, J.~A., et al.\ 2012, \mnras, 419, 3018 




\bibitem[De Lucia et al.(2006)]{de06} De Lucia, G., 
Springel, V., White, S.~D.~M., Croton, D., 
\& Kauffmann, G.\ 2006, \mnras, 366, 499 



\bibitem[Desai et al.(2007)]{de07} Desai, V., et al.\ 2007, 
\apj, 660, 1151 



\bibitem[Diaferio et al.(2001)]{di01} Diaferio, A., 
Kauffmann, G., Balogh, M.~L., White, S.~D.~M., Schade, D., 
\& Ellingson, E.\ 2001, \mnras, 323, 999 

\bibitem[Elston(1998)]{E98} Elston, R.\ 1998, \procspie, 
3354, 404 



\bibitem[Ettori et al.(2004)]{et04}
Ettori, S.~et al.\ 2004, MNRAS, 417, 13

\bibitem[Faber et al.(2007)]{fa07} Faber, S.~M., et al.\ 
2007, \apj, 665, 265 

\bibitem[Fan et al.(2008)]{2008ApJ...689L.101F} Fan, L., Lapi, A., De 
Zotti, G., \& Danese, L.\ 2008, \apjl, 689, L101 



\bibitem[Fazio et al.(1998)]{fa98} Fazio, G.~G., et al.\ 
1998, \procspie, 3354, 1024 

\bibitem[Finoguenov et al.(2007)]{fi07} Finoguenov, A., et 
al.\ 2007, \apjs, 172, 182 


\bibitem[Gal et al.(2008)]{ga08} Gal, R.~R., Lemaux, B.~C., 
Lubin, L.~M., Kocevski, D., \& Squires, G.~K.\ 2008, \apj, 684, 933 


\bibitem[Gallazzi et al. (2007)]{gal07}
Gallazzi, A.,  Charlot, S., Brinchmann, J., White, S.D.M. 2006, MNRAS, 370, 1106

\bibitem[Gehrels, N. 2006, ApJ, 303, 336]{ge06}
Gehrels, N. 2006,\apj, 303, 336

\bibitem[Geller \& Huchra(1983)]{ge83} Geller, M.~J., \& Huchra, J.~P.\ 1983, \apjs, 52, 61 


\bibitem[George et al.(2011)]{geo11} George, M. et al.\ 2011, ApJ,
  742, 125

\bibitem[Gobat et  al.(2011)]{go11} Gobat, R., et al.\ 2011, \aap, 526, A133 


\bibitem[Holden et al.(2001)]{ho11} Holden, B.~P., et al.\ 
2001, \aj, 122, 629 

\bibitem[Holden et al.(2007)]{ho07} Holden, B.~P., et al.\ 
2007, \apj, 670, 190 

\bibitem[Huertas-Company et 
al.(2008)]{2008A&A...478..971H} Huertas-Company, M., Rouan, D., Tasca, L., Soucail, G., \& Le F{\`e}vre, O.\ 2008, \aap, 478, 971 



\bibitem[Huertas-Company et 
al.(2009)]{2009A&A...497..743H} Huertas-Company, M., Tasca, L., Rouan, D., et al.\ 2009, \aap, 497, 743 


\bibitem[Huertas-Company et 
al.(2011)]{2011A&A...525A.157H} Huertas-Company, M., Aguerri, J.~A.~L., Bernardi, M., Mei, S., \& S{\'a}nchez Almeida, J.\ 2011, \aap, 525, A157 



\bibitem[Ilbert et al.(2006)]{il06} Ilbert, O., et al.\ 2006, \aap, 457, 841 

\bibitem[Ilbert et al.(2010)]{il10} Ilbert, O., et al.\ 
2010, \apj, 709, 644 

\bibitem[Jansen et 
al.(2001)]{ja01} Jansen, F., et al.\ 2001, \aap, 365, L1 



\bibitem[Jee et al.(2006)]{jee06} Jee, M.~J., White, R.~L., 
Ford, H.~C., Illingworth, G.~D., Blakeslee, J.~P., Holden, B., 
\& Mei, S.\ 2006, \apj, 642, 720 


\bibitem[Just et al.(2010)]{ju10} Just, D.~W., Zaritsky, D., 
Sand, D.~J., Desai, V., \& Rudnick, G.\ 2010, \apj, 711, 192 

\bibitem[Kashikawa et al.(2002)]{pa02} Kashikawa, N., et 
al.\ 2002, \pasj, 54, 819 


\bibitem[Kauffman and Charlot (1998)]{kau98}
Kauffman, G. \& Charlot,S. 1998, MNRAS, 294, 705


\bibitem[Kells et al.(1998)]{ke98} Kells, W., Dressler, A., 
Sivaramakrishnan, A., Carr, D., Koch, E., Epps, H., Hilyard, D., 
\& Pardeilhan, G.\ 1998, \pasp, 110, 1487 

\bibitem[Kirsch et al.(2004)]{ki04} Kirsch, M.~G.~F., et 
al.\ 2004, \procspie, 5165, 85 


\bibitem[Kodama and Arimoto (1997)]{kod97}
Kodama, T. \& Arimoto, N. 1997, AJ, 320, 41

\bibitem[Komatsu et al. (2011)]{ko11} Komatsu et al. 2011, ApJS, 192, 18



\bibitem[Kuntz \& Snowden(2008)]{ku08} Kuntz, K.~D., \& Snowden, S.~L.\ 2008, \aap, 478, 575 



\bibitem[Lubin et al.(2009)]{lu09} Lubin, L.~M., Gal, R.~R., 
Lemaux, B.~C., Kocevski, D.~D., \& Squires, G.~K.\ 2009, \aj, 137, 4867 


\bibitem[Maraston(2005)]{ma05} Maraston, C.\ 2005, \mnras, 
362, 799 



\bibitem[Markevitch et al.(1998)]{ma98} Markevitch, M., 
Forman, W.~R., Sarazin, C.~L., \& Vikhlinin, A.\ 1998, \apj, 503, 77 



\bibitem[Marleau \& Simard (1998)]{ms98}
Marleau, F.R. \& Simard, L. 1998, ApJ, 507, 585


\bibitem[McGee et al.(2009)]{mcg09} McGee, S.~L., Balogh, 
M.~L., Bower, R.~G., Font, A.~S., 
\& McCarthy, I.~G.\ 2009, \mnras, 400, 937 



\bibitem[Mei et al. (2006a)]{mei06a}
Mei, S. et al., 2006a, ApJ, 639, 81



\bibitem[Mei et al. (2006b)]{mei06b}
Mei, S. et al., 2006b, ApJ, 644, 759

\bibitem[Mei et al. (2009)]{mei09}
Mei, S. et al., 2009, ApJ, 690, 42

\bibitem[Menanteau et al. (2006)]{men06}
Menanteau, F. et al., 2006, AJ, 131, 208

\bibitem[Moran et al.(2007)]{mo07} Moran, S.~M., Ellis, 
R.~S., Treu, T., Smith, G.~P., Rich, R.~M., 
\& Smail, I.\ 2007, \apj, 671, 1503 

\bibitem[Naab et al.(2009)]{na09} Naab, T., Johansson, 
P.~H., \& Ostriker, J.~P.\ 2009, \apjl, 699, L178 


\bibitem[Nakata et al. (2005)]{na05}
Nakata, F. et al. 2005, MNRAS, 357, 1357



\bibitem[Oke et al. (1995)]{oke95}
Oke, J.~B., et al.  1995, PASP, 107, 375

\bibitem[Papovich et al.(2011)]{2011arXiv1110.3794P} Papovich, C., Bassett, 
R., Lotz, J.~M., et al.\ 2011, arXiv:1110.3794 


\bibitem[Peng, Ho, Impey, \& Rix(2002)]{pe02} 
Peng, C.~Y., Ho, L.~C., Impey, C.~D., \& Rix, H.\ 2002, \aj, 124, 266 

\bibitem[Poggianti et al.(2006)]{Po06} Poggianti, B.~M., et 
al.\ 2006, \apj, 642, 188 



\bibitem[Poggianti et al.(2009)]{po09} Poggianti, B.~M., et 
al.\ 2009, \apjl, 697, L137 



\bibitem[Poggianti et al.(2010)]{po10} Poggianti, B.~M., De 
Lucia, G., Varela, J., Aragon-Salamanca, A., Finn, R., Desai, V., von der 
Linden, A., \& White, S.~D.~M.\ 2010, \mnras, 405, 995 



\bibitem[Postman et al. (2005)]{po05}
Postman, M. et al., 2005, ApJ, 623, 721


\bibitem[Press et al. (1992)]{press2}
 Press, W.H. et al. 1992,
{\it Numerical Recipes},  Cambridge University Press, New York



\bibitem[Raichoor et al.(2011)]{ra11} Raichoor, A., et al.\ 
2011, \apj, 732, 12 


\bibitem[Raichoor et al.(2012)]{rau12} Raichoor, A., et al.\ 
2012, \apj, 745, 130

\bibitem[Rettura et al. (2006)]{re06}
Rettura, A. et al., 2006, A\&A, 458, 717


\bibitem[Rettura et al.(2011)]{re11} Rettura, A., et al.\ 
2011, \apj, 732, 94 


\bibitem[Rosati et al.\ (1999)]{ro1999} 
Rosati, P., Stanford, S.A., Eisenhardt, P.R., Elston, R., Spinrad, H., Stern, D., Dey, A. 1999, ApJ, 118, 76

\bibitem[Saxton et al.(2005)]{sax05} Saxton, R.~D., Altieri, 
B., Read, A.~M., Freyberg, M.~J., Esquej, M.~P., 
\& Bermejo, D.\ 2005, \procspie, 5898, 73 


\bibitem[Shankar et al.(2011)]{sha11} Shankar, F., Marulli, 
F., Bernardi, M., Mei, S., Meert, A., \& Vikram, V.\ 2011, arXiv:1105.6043 


\bibitem[Schlegel et al. (1998)]{SFD98}
Schlegel, D. J., Finkbeiner, D. P., \& Davis, M. 1998, \apj, {500}, 525



\bibitem[Scodeggio (2001)]{sco01}
Scodeggio, M. 2001, AJ, 121, 241


\bibitem[Shen et al.(2003)]{she03} Shen, S., Mo, H.~J., 
White, S.~D.~M., Blanton, M.~R., Kauffmann, G., Voges, W., Brinkmann, J., 
\& Csabai, I.\ 2003, \mnras, 343, 978 


\bibitem[Smith et al. (2005)]{smi05}
Smith, G.P. et al. 2005, ApJ, 620, 78


\bibitem[Sirianni et al. (2005)]{si}
Sirianni, M. et al. 2005, PASP, 117, 1049




\bibitem[Stanford et al. (1997)]{stan97}
Stanford, S. A., Elston, R., Eisenhardt, P. R., Spinrad, H., Stern, D., Dey, A. 1997, AJ, 114, 2232

\bibitem[Stanford et al. (2001)]{stan01}
Stanford, S. A., Holden, B., Rosati, P., Tozzi, P., Borgani, S., Eisenhardt, P. R., Spinrad, H. 2001, 552, 502

\bibitem[Strazzullo et al.(2010)]{stra10} 
Strazzullo, V., et al.\ 2010, \aap, 524, A17 



\bibitem[Str{\"u}der et  al.(2001)]{stru10} 
Str{\"u}der, L., et al.\ 2001, \aap, 365, L18 


\bibitem[Tanaka et al.(2007)]{ta07} Tanaka, M., Kodama, T., 
Kajisawa, M., Bower, R., Demarco, R., Finoguenov, A., Lidman, C., \& Rosati, P.\ 2007, \mnras, 377, 1206 


\bibitem[Tanaka et al.(2009)]{ta09} 
Tanaka, M., Lidman, C., Bower, R.~G., Demarco, R., Finoguenov, A., Kodama, T., Nakata, F., \& Rosati, P.\ 2009, \aap, 507, 671 

\bibitem[Tanaka et al.(2010)]{Ta10} 
Tanaka, M., Finoguenov, A., \& Ueda, Y.\ 2010, \apjl, 716, L152 



\bibitem[Treu et al.(2003)]{tr03} 
Treu, T., Ellis, R.~S., Kneib, J.-P., Dressler, A., Smail, I., Czoske, O., Oemler, A., \& Natarajan, P.\ 2003, \apj, 591, 53 

\bibitem[Turner et al.(2001)]{tu01} 
Turner, M.~J.~L., et al.\ 2001, \aap, 365, L27 

\bibitem[Valentinuzzi et al.(2010)]{va10} Valentinuzzi, T., 
et al.\ 2010, \apj, 712, 226 



\bibitem[van Dokkum et al.(1998)]{V98} van Dokkum, P.~G., 
Franx, M., Kelson, D.~D., Illingworth, G.~D., Fisher, D., \& Fabricant, D.\ 1998, \apj, 500, 714


\bibitem[van Dokkum \& Franx (2001)]{vDf01}
van Dokkum, P.~G. \& Franx, M.\ 2001, \apj, 553, 90

\bibitem[Watson et al.(2001)]{wa01} 
Watson, M.~G., et al.\ 2001, \aap, 365, L51 

\bibitem[Wilman et al.(2009)]{wi09} Wilman, D.~J., Oemler, 
A., Jr., Mulchaey, J.~S., McGee, S.~L., Balogh, M.~L., \& Bower, R.~G.\ 2009, \apj, 692, 298 



\bibitem[Yamada et al. (2002)]{yam02}
Yamada, T., Koyama, Y., Nakata, F., Kajisawa, M., Tanaka, I., Kodama, T., Okamura, S., De Propris, R. 2002, ApJL, 577, 89

\bibitem[York et al.(2000)]{yo00} 
York, D.~G., et al.\ 2000, \aj, 120, 1579 



\bibitem[Zhang et al.(2004)]{zha04} 
Zhang, Y.-Y., Finoguenov, A., B{\"o}hringer, H., Ikebe, Y., Matsushita, K., \& Schuecker, P.\ 2004, \aap, 413, 49 


\end{thebibliography}
\end{document}